\pdfoutput=1
\documentclass{aa}
\usepackage[varg]{txfonts}
\usepackage{graphicx}
\usepackage{t1enc}
\newcommand{\F}[1]   {\ensuremath{\mathbf{F}[#1]}}
\newcommand{\Fp}[1]  {\ensuremath{\mathbf{F'}[#1]}}
\newcommand{\kms}    {km~s$^{-1}$}
\newcommand{\RA}[4]  {$\alpha(J2000)=#1^\mathrm{h}#2^\mathrm{m}#3\fs#4$}
\newcommand{\DEC}[4] {$\delta(J2000)=#1\degr#2'#3\farcs#4$}
\newcommand{\nh}     {NH$_3$}
\newcommand{\Msol}   {$M_{\sun}$}
\newcommand{\Lsol}   {$L_{\sun}$}

\begin{document}

\title{
Analysis and test of the central-blue-spot infall hallmark
}
\titlerunning{Central-blue-spot infall hallmark}

\author{
R. Estalella\inst{1} 
\and
G. Anglada\inst{2} 
\and
A. K. D\'{\i}az-Rodr\'{\i}guez\inst{2} 
\and
J. M. Mayen-Gijon\inst{2}
}
\institute{
Departament de F\'{\i}sica Qu\`antica i Astrof\'{\i}sica,
Institut de Ci\`encies del Cosmos, Universitat de Barcelona, IEEC-UB,
Mart\'{i} i Franqu\`es, 1, E-08028 Barcelona, Spain. 
\and 
Instituto de Astrof\'{\i}sica de Andaluc\'{\i}a, CSIC, 
Glorieta de la Astronom\'{\i}a, s/n, E-18008 Granada, Spain.
}
\authorrunning{Estalella et al.}

\date{
Received 2018 December 31/
Accepted 2019 April 24
}

\abstract{}{
The infall of material onto a protostar, in the case of optically thick line
emission, produces an asymmetry in the blue- and red-wing line emission. For an
angularly resolved emission, this translates in a blue central spot in the
first-order moment (intensity weighted velocity) map. 
}{
An analytical expression for the first-order moment intensity as a function of
the projected distance was derived, for the cases of infinite and finite
infall radius. The effect of a finite angular resolution, which requires
the numerical convolution with the beam, was also studied. 
}{ 
This method was applied to existing data of several star-forming regions, 
namely G31.41+0.31 HMC, B335, and LDN 1287,
obtaining good fits to the first-order moment intensity maps, and deriving
values of the central masses onto which the infall is taking place
(G31.41+0.31 HMC: 70--120 \Msol{}; 
B335: 0.1 \Msol{}; 
Guitar Core of LDN 1287: 4.8 \Msol{}). 
The central-blue-spot infall hallmark appears to be a robust and reliable
indicator of infall.
}{}

\keywords{
ISM: jets and outflows -- 
ISM: individual objects: G31.41+0.31 HMC, B335, LDN 1287 --
stars: formation 
}

\maketitle

\section{Introduction}

The early phases of star formation are characterized by infall motions of
ambient material onto a central protostellar object. However, obtaining
unequivocal observational evidence for these motions constitutes a long-standing
problem. Rotation, infall and outflow motions can be present simultaneously in
the early phases of star formation, and may produce similar observational
features in the line profiles, making difficult an unambiguous interpretation of
the observations.

The most common method used so far to search for evidence of infall is  based on
the so-called ``blue asymmetry''. This signature consists in  the appearance of
two peaks in the spectral line profiles, with the  blueshifted peak stronger
than the redshifted peak 
\citep[e.g.][]{Zho94, Kla07, Wu07}.
However, the effect of protostellar infall on molecular  line profiles cannot be
easily isolated from those of other dynamical  processes, resulting in
ambiguities 
\citep[e.g.][]{Pur06, Szy07}.
Inverse P-Cygni profiles, which consist in the detection  of absorption at
redshifted velocities against a bright background continuum  source, have been
observed in molecular lines against a bright  background HII region 
\citep[e.g.][]{Ket87, Zha98}
or against the  bright dust continuum emission of the hot protostellar core
itself  
\citep[e.g.][]{DiF01, Gir09},
and have been interpreted as infall motions of the surrounding envelope.
Nevertheless,  it uniquely indicates that foreground matter is moving towards a
hotter  source, no matter how or if it is indeed gravitationally bound. See 
\citet{May14} and \citet{May15} for a thorough review of infall  signatures.

\begin{figure}[htb]
\centering
\resizebox{\hsize}{!}{\includegraphics{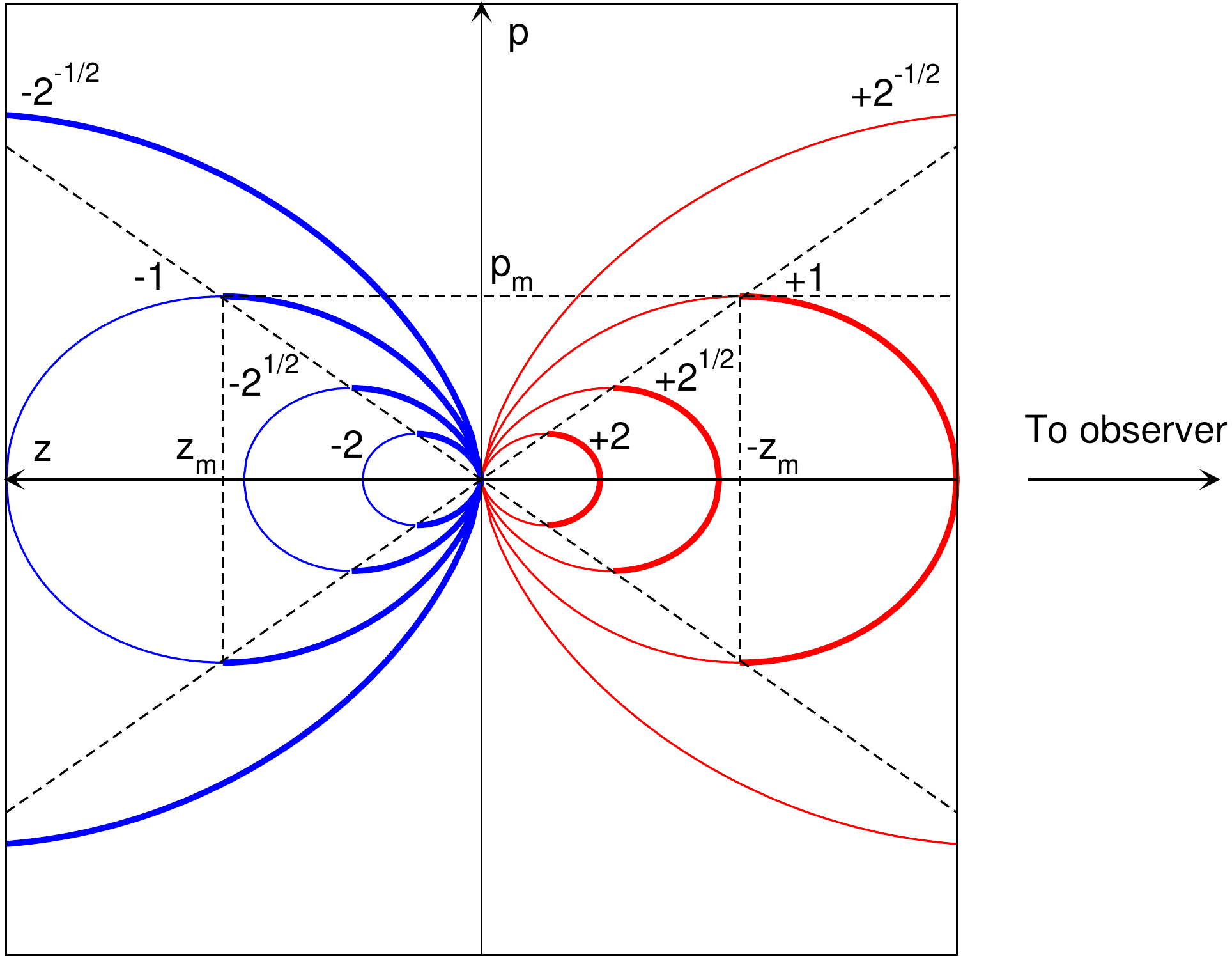}}
\caption{\label{feggs}
Intersection of the surfaces of equal line-of-sight (LOS) velocity, $v_z$, for a
collapsing protostellar core, with the plane $(p, z)$ (see \S \ref{Iprofile}).
The LOS-velocities are different pairs of negative and positive velocities 
($v_z=\pm2^{-1/2}, \pm1, \pm2^{1/2}, \pm2$). 
The blue lines are the contours for $v_z<0$, and the
red lines for $v_z>0$.  
The observer, located to the right, at $p=0$, $z=-\infty$, sees the emission
coming from the part of the surfaces of equal LOS-velocity facing the observer,
traced with thick lines.
The dashed lines indicate the position of $(z_m, p_m)$ for each contour, where
$p_m$ is the maximum value of $p$, and $z_m$ the corresponding value of $z$. 
The square frame is drawn at $p=\pm1$, $z=\pm1$.
}
\end{figure}

Anglada et al. (1991) introduce a more complete  signature, the
``3D spectral imaging infall signature'', which is based  on the spatial
distribution of the line emission intensity in the images  as a function of the
line-of-sight (LOS) velocity (i.e.\ the channel  maps). This signature,
appropriate for angularly resolved sources,  results as an extension of the
formalism initially developed by  \citet{Ang87} for the study of an angularly
unresolved infalling core.  These signatures are focused on relatively high
velocities, in order to  avoid confusion with emission from the ambient cloud at
low velocities.  With the assumption of gravitational infall motions dominating
the  kinematics over turbulent and thermal motions, and spherical symmetry  with
an infall velocity increasing inwards, it can be shown that the  points of the
infalling core with the same LOS velocity form closed  surfaces. The
equal-LOS-velocity surfaces are a nested set of surfaces  with the same shape,
decreasing in size with increasing absolute value  of the LOS velocity, and
converging to the position of the core center  (see Fig.\ \ref{feggs}).

\begin{figure}[htb]
\centering
\resizebox{\hsize}{!}{\includegraphics{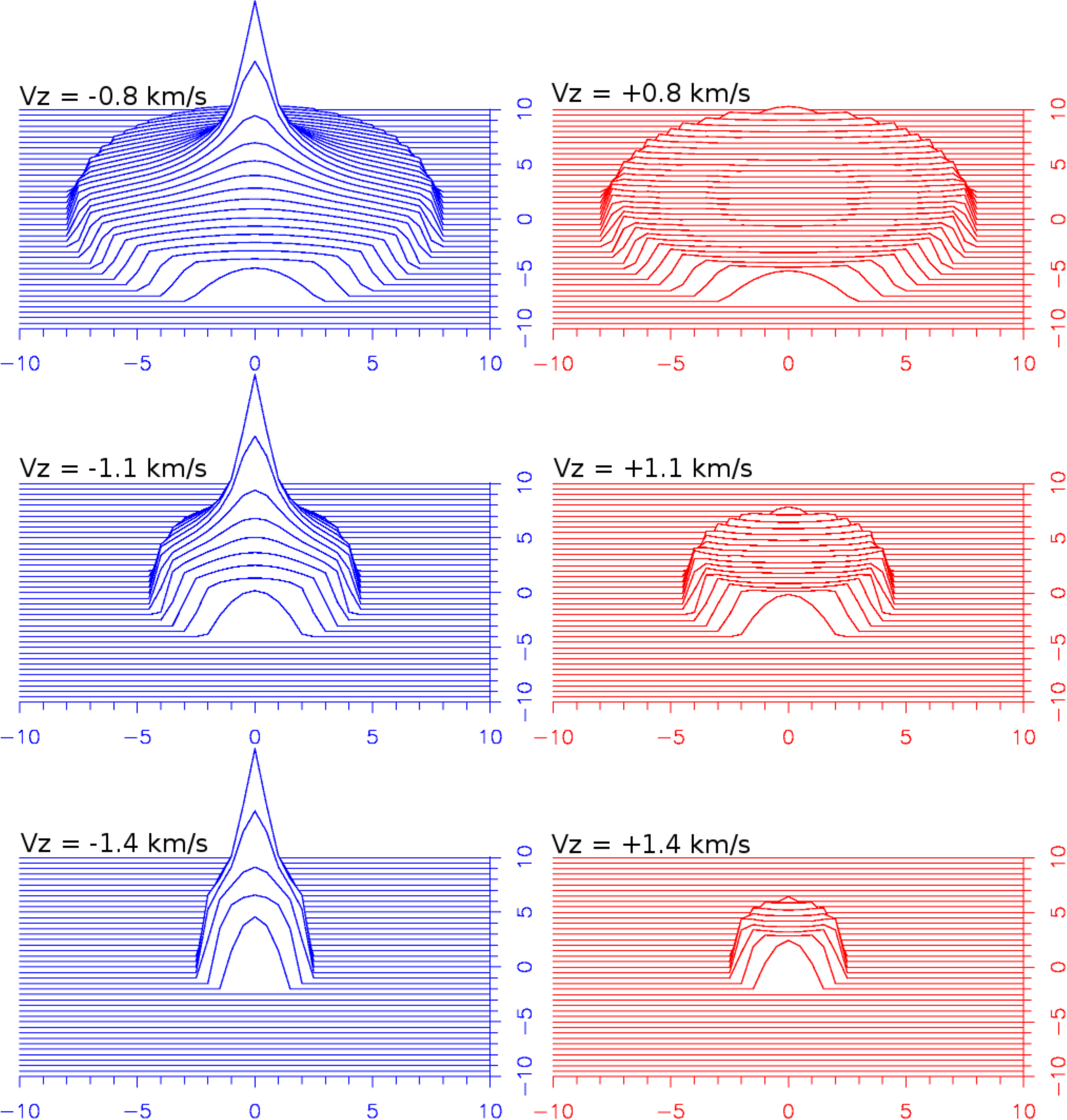}}
\caption{\label{fintensity}
Intensity maps for pairs of LOS-velocities, 
$V_z=\pm0.8$~\kms\ (top),
$\pm1.1$~\kms\ (middle), and
$\pm1.4$~\kms\ (bottom), calculated for a 
a mass $M_\ast=1$~\Msol, 
a distance of 140~pc, and 
a beam of $0\farcs1$. The axes are position offsets labeled in arcsec.
The intensity scale is the same for all maps.
Note the sharp peak at the center of the blueshifted LOS-velocity maps (left),
and the flat, slightly concave shape of the redshifted LOS-velocity maps
(right), and the more compact emission for high values $|V_z|$ (bottom) than
for low values of $|V_z|$ (top).
}
\end{figure}

Because the equal-LOS-velocity surfaces are closed surfaces, a given LOS, in
general, intersects the same surface twice. However, if the opacity is high
enough, only a narrow layer at the front side of the  equal-LOS-velocity
surface is observable. Hence, the intensity map at a given LOS velocity is an
image of the excitation temperature distribution in the side of the
equal-LOS-velocity surface facing the  observer (thick lines in Fig.\
\ref{feggs}), while the  emission from the rear side remains hidden (thin lines
in Fig.\  \ref{feggs}). 
Thus, the shapes of the blueshifted and redshifted emitting
regions\footnote{Throughout the  paper, the LOS velocity considered is relative
to the systemic velocity of the source.} are different. Since the temperature
increases inwards, the integrated blueshifted emission comes from a region
closer, on average, to the central protostar (and, therefore, hotter) than the
corresponding  redshifted emission, resulting in asymmetric line profiles, with
the blueshifted wing stronger than the redshifted wing \citep{Ang87}. The 
difference is still more remarkable for a pair of blueshifted--redshifted
channel maps. Assuming that the maps are centered on the position of the
protostar, for the redshifted channel the  intensity is slightly lower at
the center of the map than at the edges, because towards the center the side of
the equal-LOS-velocity surface facing the observer is slightly farther away
from the center of the core and thus, colder than the rest of the surface. For
the blueshifted  channel, the intensity increases sharply towards the center of
the image since this emission comes from a region very close to the center of
the  core and, therefore, very hot (see Figs.\ \ref{fintensity}, 
\ref{fradialprofile} and \ref{fradialprofile2}). In addition, since the  size of
the equal-LOS-velocity surfaces decreases with increasing  absolute value of the
LOS velocity, the emission becomes more compact  for increasing absolute value
of the LOS velocity (see Figs.\  \ref{fintensity} and \ref{fradialprofile2}).

This behavior of the intensity maps for redshifted and blueshifted LOS
velocities produces a characteristic signature in the intensity-weighted mean
velocity (first-order moment) map, which was pointed out by \citet{May14}, 
i.e.\ that the central region of the first-order map appears blueshifted because
of the higher weight of the strong blueshifted emission. Additionally,  the
integrated intensity, (zeroth-order moment) peaks towards the central position.
At larger distances from the center, the integrated intensity decreases, the
blue and redshifted intensities become similar, and the intensity-weighted mean
velocity approaches the systemic velocity of the cloud. Therefore, the
first-order moment of an infalling envelope is characterized by a compact spot
of blueshifted emission towards the position of the zeroth-order moment peak.
This infall hallmark is designated as the ``central blue spot'' \citep{May14,
May15}. One of the advantages of the central-blue-spot infall hallmark is that
its detection does not require a beforehand knowledge of the systemic velocity
of the cloud. An accurate knowledge of the systemic velocity is critical in
searching for infall through the analysis of asymmetries in the line profiles. 

\citet{May14} and \citet{May15} identify both the 3D spectral imaging
infall signature and the central-blue-spot hallmark in high-angular resolution
maps of the emission of several \nh{} transitions towards G31.41+0.31 HMC, and
compare the observed emission with the predictions of a spherically
symmetric model with full transport of radiation calculation
\citep{Oso09}.
 
If there is rotation, the LOS velocity has contributions from both the infalling
velocity and the rotation velocity. \citet{May14} and \citet{May15}, in the
analysis of the \nh{} data of G31.41+0.31 HMC, discuss qualitatively how
rotation affects the channel maps of an infalling core. These authors find
that the radial intensity profile of the image for a given LOS-velocity channel
is stretched towards the side where rotation has the same sign than the channel
velocity, and it is shrunk on the opposite side. Nevertheless, as in the
non-rotating case, the images in blue-shifted channels present a centrally
peaked intensity distribution, while in the red-shifted channels they present a
flatter intensity distribution. Thus, the rotation signature makes the  spatial
intensity profiles asymmetric with respect to the central position but it does
not mask the 3D spectral imaging infall signature of \citet{Ang91}.

Regarding the first-order moment map, \citet{May15} explores how the
central-blue-spot hallmark of an infalling core is modified by the presence of
rotation. He finds that rotation makes the central-blue-spot even bluer
and moves it off the center towards the half of the core where rotation tends
to shift velocities to the blue. Additionally, a dimmer red spot appears
symmetrically located on the opposite side of the rotation axis. 

In the present paper we studied quantitatively the central-blue-spot infall
hallmark,  restricted to the spherically symmetric case without rotation, 
taking as a basis the work of \citet{Ang87, Ang91}.  Using the same assumptions
than in these papers, we derived  analytical expressions for the intensity
profiles (\S 2),  the line profiles (\S 3), and  the first-order moment of the
intensity profile (\S 4)  as functions of the angular distance from the center. 
This was done for the case of an infinite infall radius, without considering the
effect of a finite angular resolution. 
The details of the derivation for arbitrary values of the power-law indices are
given in Appendix \ref{ap_infinite}.  
The effect of a finite spectral resolution is addressed in \S 4.2 and Appendix
\ref{ap_spectralresolution}, while 
the effect of a finite angular resolution is studied in \S 4.3 and Appendix
\ref{ap_convolution}. 
The case of a finite infall radius is presented in \S 5,
where an analytical expression for the first-order moment is obtained. 
The details of the derivation for arbitrary values of the power-law indices are
given in Appendix \ref{ap_finite}. 
The transformation between reduced units and practical units is described in \S
6. 
The results are applied to several cases (G31.41+0.31 HMC, B335, LDN 1287) in \S
7, with the analysis of pre-existing data that show the central-blue-spot infall
hallmark. 
Finally, the conclusions are given in \S 8.

\section{Intensity profile}
\label{Iprofile}

Based on \citet{Ang87, Ang91}, we are assuming an infalling molecular gas core,
with infall velocity and temperature given by power laws of the radius
\citep[Eq.\ 1 of ][]{Ang87},
\begin{eqnarray}\label{eq1} 
V/V_0 &=& (R/R_0)^{-\alpha}, \nonumber\\ 
T/T_0 &=& (R/R_0)^{-\beta},
\end{eqnarray}
where $R_0$ is a reference radius.
The power-law indices of the infall velocity and temperature are taken with a
value $\alpha=\beta=1/2$, i.e.\ 
free-fall velocity, and optically thin dust heating from a central protostar, 
which are characteristic of the main accretion phase in the Larson
collapse model \citep{Lar72} or in the \citet{Shu77} inside-out collapse model.  
In the Appendices \ref{ap_infinite} and \ref{ap_finite}
the case of arbitrary values of the power-law indices is developed.

Let the coordinate $z$ be along the line of sight, positive outwards from the
observer, and $p$ the impact parameter, i.e.\ the distance to the center
projected on the plane of the sky
(see Fig.\ \ref{feggs}). 
All coordinates are lengths in units of $R_0$, so that the distance to the
center in units of $R_0$ is
\begin{equation}
r\equiv R/R_0= (p^2+z^2)^{1/2}.
\end{equation}
Let us define the reduced line-of-sight (LOS) velocity and temperature
\begin{eqnarray}\label{eq2} 
v_z &\equiv& V_z/V_0, \nonumber\\ 
t   &\equiv& T/T_0.
\end{eqnarray}

\begin{figure}[htb]
\centering
\resizebox{\hsize}{!}{\includegraphics{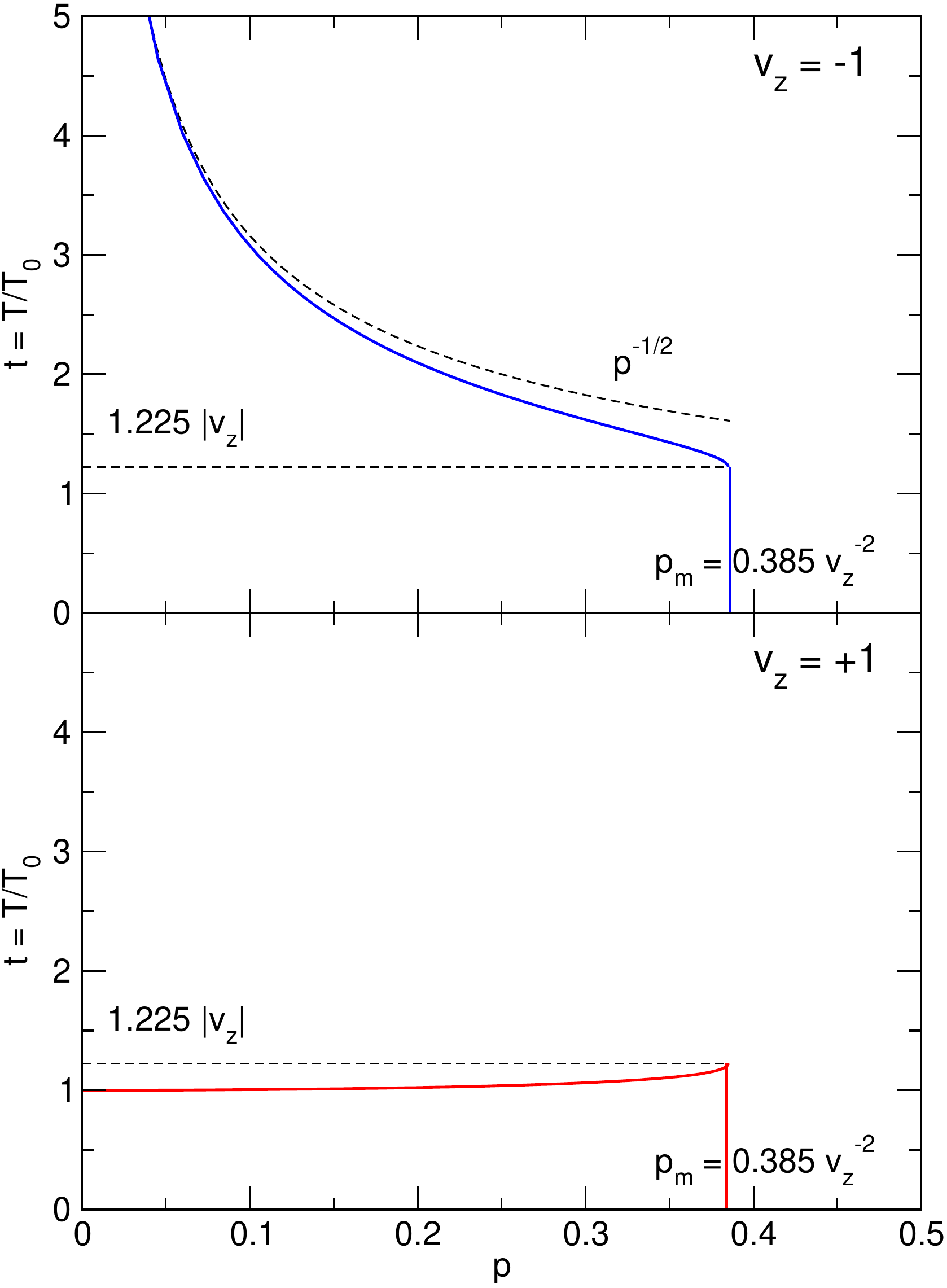}}
\caption{\label{fradialprofile}
Intensity profiles $t(p)$ for 
$v_z=-1$ (top panel, blue) and 
$v_z=+1$ (bottom panel, red).  
The different dashed lines indicate the blue and red wing temperature at the
maximum value of $p$, $p=p_m$, and  
the blue wing asymptotic behavior for $p\ll1$. 
} 
\end{figure}

\begin{figure}[htb]
\centering
\resizebox{\hsize}{!}{\includegraphics{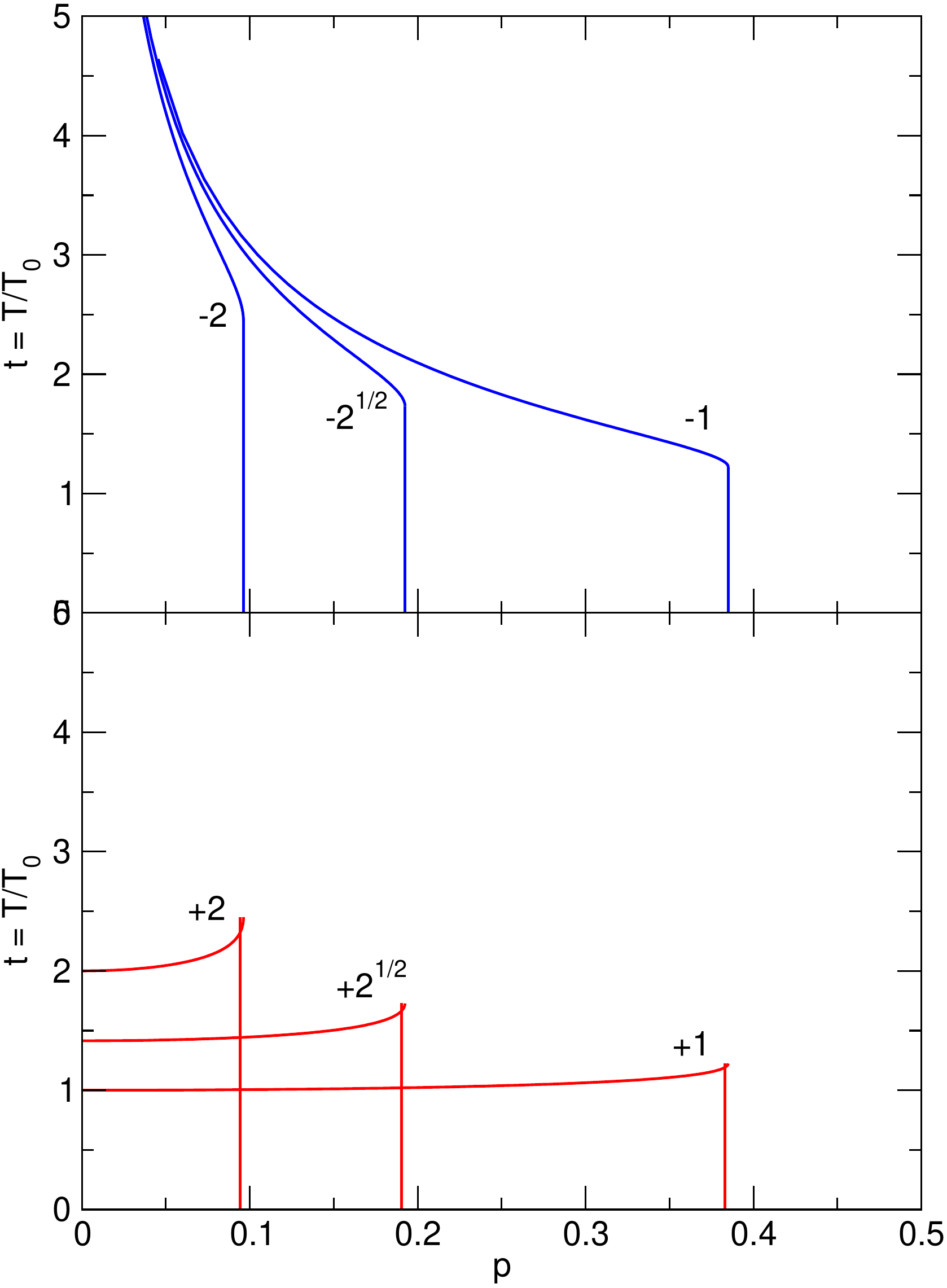}}
\caption{\label{fradialprofile2}
Intensity profiles $t(p)$ for pairs of positive and negative
velocities, $v_z=\pm1, \pm2^{1/2}, \pm2$.
}
\end{figure}

Equations 9, 10 of \citet{Ang91} can be expressed in reduced variables as
\begin{eqnarray}\label{etp}
p &=& \left[ |z/v_z|^{4/3} - z^2\right]^{1/2}, \nonumber\\
t &=& |v_z/z|^{1/3}.
\end{eqnarray}
For a given $|v_z|$ Eq.\ \ref{etp} gives the parametric equation (with $z$ as
parameter) of the intensity profile, $t(p)$. 
We are assuming that the intensity observed at a given LOS velocity comes from
the part of the equal-LOS-velocity surface facing the observer.  
Thus, the blue-wing intensity profile ($v_z<0$) is obtained for $0<z<z_m$, while 
the red-wing intensity profile ($v_z>0$) is obtained for $-z^\ast<z<-z_m$, 
where $z_m$ is the value of $z$ for which $p$ is maximum for a given $v_z$,
$p=p_m$ (see Fig.\ \ref{feggs}), and
$z^\ast$ is the maximum value of $z$ for a given equal-LOS-velocity surface.
The value of $z^\ast$ is obtained from Eq.\ \ref{etp} for $p=0$, 
$z^\ast=|v_z|^{-2}$. 
The value of $z_m$ can be obtained from the derivative of $p(z)$ given in Eq.\
\ref{etp}, and the values obtained for $z_m$ and $p_m$ are
\begin{eqnarray}\label{eq4} 
z_m &=& (2/3)^{3/2}\, v_z^{-2}= 0.544\,v_z^{-2}, \nonumber\\ 
p_m &=& 2/3^{3/2}  \, v_z^{-2}= 0.385\,v_z^{-2}.
\end{eqnarray}
The ratio $z_m/p_m$ is independent of $v_z$, $z_m/p_m=\sqrt{2}$, meaning that
the points $(z_m, p_m)$ are aligned along a straight line passing
through the center, with a slope of $1/\sqrt{2}$ (see Fig.\ \ref{feggs}).

For any LOS-velocity the emission is confined inside a projected distance
$p<p_m$.  For blueshifted LOS-velocities ($v_z<0$) the intensity increases
sharply for small projected distances $p$, while for redshifted LOS-velocity
($v_z>0$) the intensity is almost flat up to $p_m$ (see Figs.\
\ref{fradialprofile} and \ref{fradialprofile2}). 
This can be seen in Fig.\
\ref{fintensity}, where we show the intensity maps, for LOS pairs of positive
and negative LOS velocities.

For the maximum projected distance $p_m$ the red-wing intensity is maximum, and
equal to the minimum blue-wing intensity (see Fig.\ \ref{fradialprofile}),
\begin{equation}\label{eq5}
t(p_m)=(3/2)^{1/2}\, v_z= 1.225\,v_z.
\end{equation}
The blue wing intensity for small projected distances is very high, and for the
adimensional temperature and LOS velocity $t\gg v_z$ we obtain the asymptotic
behavior (see Fig.\ \ref{fradialprofile})
\begin{equation}\label{eq6}
t(p)\simeq p^{-1/2} \quad (p\ll v_z^{-2}).
\end{equation}

\section{Line profile}
\label{Lprofile}

\begin{figure}[htb]
\centering
\resizebox{\hsize}{!}{\includegraphics{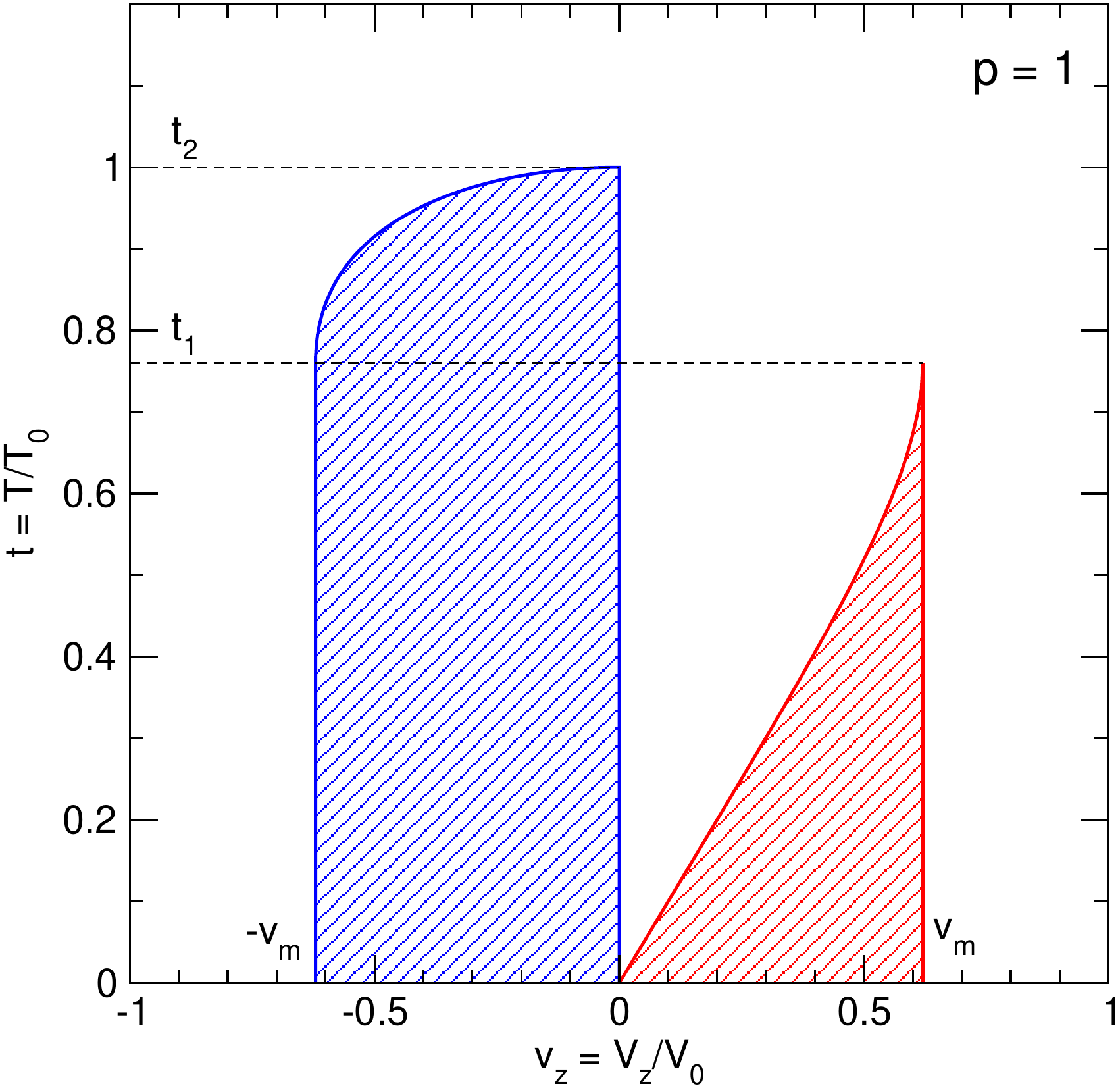}}
\caption{\label{flineprofile}
Line profile showing the emission at a projected distance
$p$ as a function of LOS velocity.
The blue wing emission encompasses LOS velocities from $-v_m$ to 0, 
and its peak value (at $v_z=0$) is $t_2=p^{-1}$.
The red wing emission encompasses LOS velocities from 0 to 
$v_m=0.620 p^{-1/2}$, and its peak value (at $v_z=v_m$) is $t_1=0.760 p^{-1}$.
}
\end{figure}

\begin{figure}[htb]
\centering
\resizebox{\hsize}{!}{\includegraphics{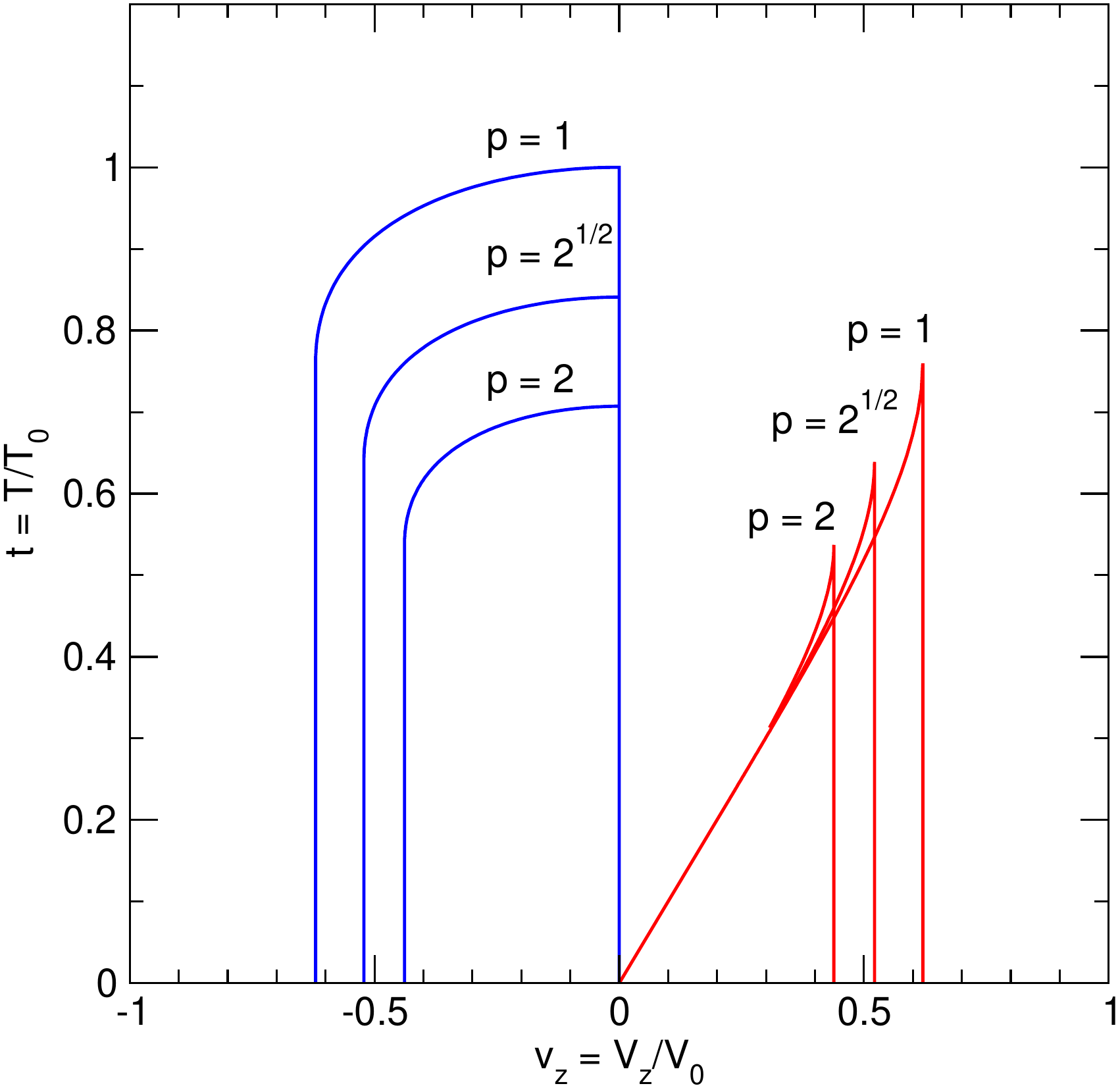}}
\caption{\label{flineprofile2}
Line profiles for values of $p= 1, 2^{1/2}, 2$.
}
\end{figure}

In order to calculate the first-order moment intensity, we need to obtain the
line intensity for any value of $p$. Thus, we want to derive the intensity and LOS
velocity for a given projected distance $p$. 
From Eq.\ \ref{etp} we can obtain easily the LOS velocity and temperature as
functions of $p$ and $z$,
\begin{eqnarray}\label{evzt}\label{eq7}
v_z &=& \frac{-z}{(p^2+z^2)^{3/4}}, \nonumber\\
t   &=& \frac{1}{(p^2+z^2)^{1/4}}.
\end{eqnarray}
These equations can be interpreted as the parametric equation (with $z$ as
parameter) of the line profile $t(v_z)$ for a given projected distance $p$ (see
Figs.\ \ref{flineprofile} and \ref{flineprofile2}). 

The line of sight with a given $p$ intersects equal-LOS-velocity surfaces 
of decreasing size, down to
that corresponding to $p_m=p$. For this equal-LOS-velocity surface, the
blueshifted velocity is minimum (maximum absolute value) and the redshifted
velocity is maximum (see Fig.\ \ref{feggs}).
This occurs at the points with coordinate $z=\pm\sqrt{2}p$.

\subsection{Blue wing}

The observable intensity at blueshifted velocities ($v_z<0$) 
is obtained for $0<z<\sqrt{2}p$. 
For a given value of $p$, the minimum blueshifted intensity
$t^\mathrm{blue}_\mathrm{min}$ and minimum velocity
$v^\mathrm{blue}_\mathrm{min}$ (maximum absolute value) are obtained at the
point  $(p, z=+\sqrt{2}p)$, 
\begin{eqnarray}\label{eq8} 
t^\mathrm{blue}_\mathrm{min} &\equiv&  t_1 = 3^{-1/4}\,p^{-1/2}=          0.760\,p^{-1/2}, \nonumber\\
v^\mathrm{blue}_\mathrm{min} &\equiv& -v_m = -2^{1/2}3^{-3/4}\,p^{-1/2}= -0.620\,p^{-1/2},
\end{eqnarray}
while the maximum blueshifted intensity $t^\mathrm{blue}_\mathrm{max}$ and maximum velocity
$v^\mathrm{blue}_\mathrm{max}=0$ are obtained at the point $(p, z=0)$
\begin{eqnarray}\label{eq9} 
t^\mathrm{blue}_\mathrm{max} &\equiv& t_2 = p^{-1/2}, \nonumber\\
v^\mathrm{blue}_\mathrm{max} &=& 0.
\end{eqnarray}

\subsection{Red wing}

The observable intensity at redshifted velocities ($v_z>0$) 
is obtained for $-\infty<z<-\sqrt{2}p$. 
For a given value of $p$, the minimum redshifted intensity $t^\mathrm{red}_\mathrm{min}=0$
and minimum velocity $v^\mathrm{red}_\mathrm{min}=0$ are obtained at the point 
$(p, z=+\infty)$, 
\begin{eqnarray}\label{eq10} 
t^\mathrm{red}_\mathrm{min} &=& 0, \nonumber\\ 
v^\mathrm{red}_\mathrm{min} &=& 0,
\end{eqnarray}
while the maximum redshifted intensity $t^\mathrm{red}_\mathrm{max}$ and maximum velocity
$v^\mathrm{red}_\mathrm{max}$ are obtained at the point $(p, z=-\sqrt{2}p)$,
\begin{eqnarray}\label{eq11} 
t^\mathrm{red}_\mathrm{max} &\equiv& t_1 = 3^{-1/4}\,p^{-1/2}=        0.760\,p^{-1/2}, \nonumber\\ 
v^\mathrm{red}_\mathrm{max} &\equiv& v_m = 2^{1/2}3^{-3/4}\,p^{-1/2}= 0.620\,p^{-1/2}.
\end{eqnarray}

\section{First-order moment for an infinite infall radius}

\subsection{Dependence on the projected distance}
\label{Mprofile}

We are interested in calculating the first-order normalized moment of the line
profile as a function of the projected distance,
\begin{equation}\label{eq12} 
\mu_1(p)= \frac{\mu'_1(p)}{\mu_0(p)},
\end{equation}
where $\mu_0$ and $\mu_1$ are the zeroth-order and unnormalized 
first-order moments,
\begin{eqnarray}\label{edefmu01}\label{eq13} 
\mu_0(p)  &=&  \int_\mathrm{line} t(p, v_z)\,dv_z, \nonumber\\ 
\mu'_1(p) &=& \int_\mathrm{line} v_zt(p, v_z)\,dv_z. 
\end{eqnarray}
The moment $\mu_0$ has units of intensity times velocity, while 
$\mu_1$ has units of velocity.

From Eq.\ \ref{evzt} we can obtain $v_z$ as an explicit function of $t$,
\begin{equation}\label{eq14}  
|v_z|= t{(1-p^2t^4)^{1/2}},
\end{equation}
where the blue-wing profile is obtained for $3^{-1/4}\,p^{-1/2}<t<p^{-1/2}$,
and the red-wing profile for $0<t<3^{-1/4}\,p^{-1/2}$.
Since we know the inverse function $v_z(t)$, we can evaluate the integrals by
integration by parts,
\begin{eqnarray} 
\mu_0  &=& \int t\,dv_z= v_z t - \int v_z\,dt, \nonumber\\ 
\mu'_1 &=& \int v_zt\,dv_z= 
\frac{1}{2}v_z^2 t -\frac{1}{2}\int v_z^2\,dt.
\end{eqnarray}

\begin{figure}[htb]
\centering
\resizebox{\hsize}{!}{\includegraphics{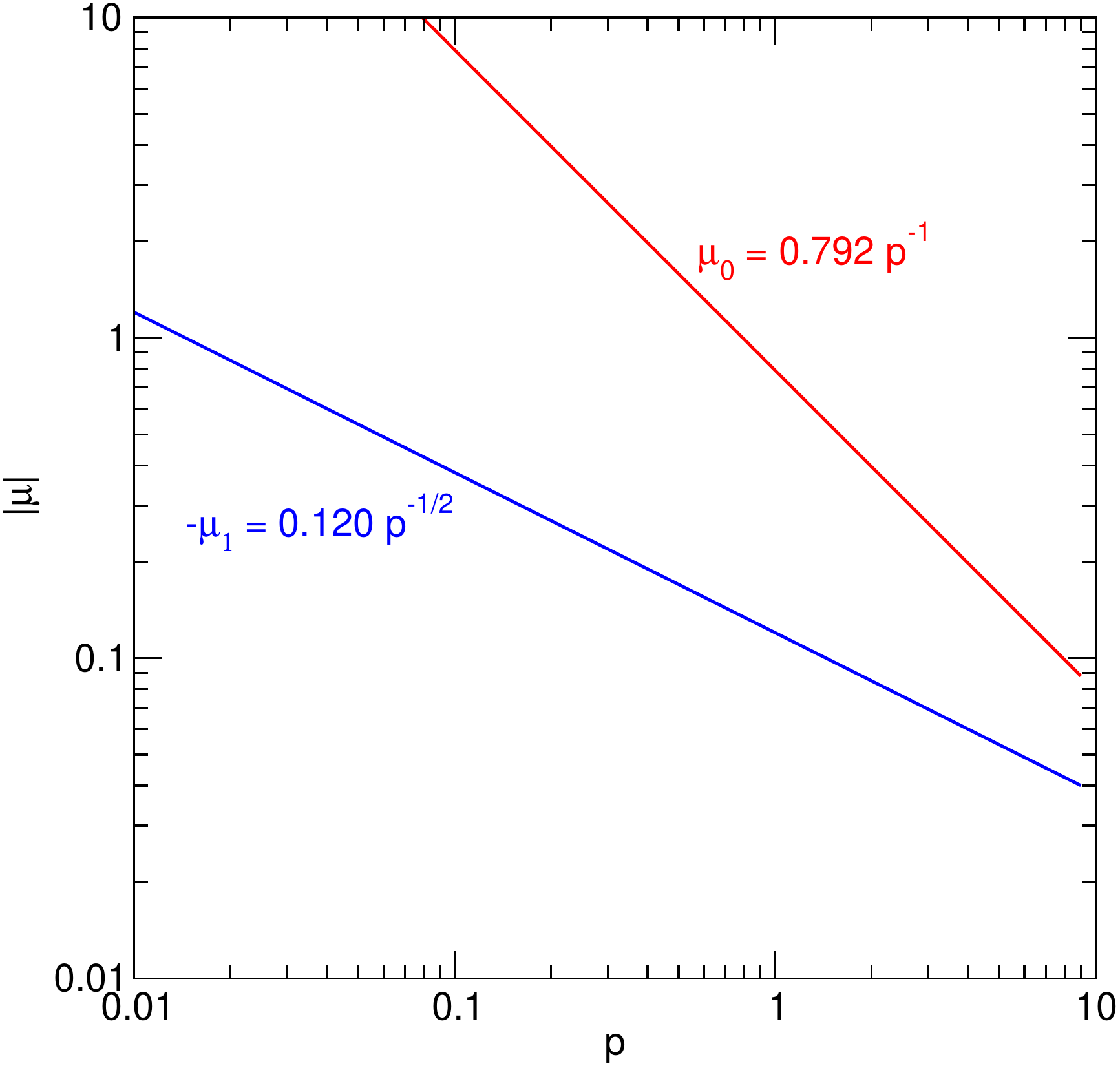}}
\caption{\label{fmu01}
Log-log plot of the moments $\mu_0$ and $-\mu_1$, as a function of
projected distance $p$. }
\end{figure}

The details of the calculation of these integrals are given in
Appendix \ref{ap_infinite}.
As described in the appendix, 
the final results obtained can be summarized as follows. 
The moments $\mu_0$, $\mu'_1$, and $\mu_1$ are power laws of the
projected distance $p$, and given by
\begin{eqnarray}\label{eqresult1}
\mu_0(p)  &=& H_0\,p^{-1}, \nonumber\\
\mu'_1(p) &=& H_1\,p^{-3/2},\\
\mu_1(p)  &=& [H_1/H_0]\,p^{-1/2}, \nonumber
\end{eqnarray}
with $H_0$ and $H_1$ given by
\begin{eqnarray}\label{eqresult2}
H_0 &=& \frac{\sqrt{2}}{2}+\frac{\pi}{8}-\frac{1}{2}\arcsin{\frac{1}{\sqrt{3}}}
= 0.792,\nonumber\\
H_1 &=& -\frac{2}{21}= -0.095, \\
H_1/H_0 &=& -0.120. \nonumber
\end{eqnarray}
A plot of the zeroth- and first-order moments can be seen in Fig.\ \ref{fmu01}.

\subsection{Effect of a finite spectral resolution}

Let us assume that we are observing the lines with an spectrometer with a finite
spectral resolution. The response of the spectrometer can be represented by the
convolution of the ``real'' line profile with the instrumental response, which
has a width equal to the spectral resolution. In general, we can assume that 
instrumental response is symmetric with respect to velocity. As shown in
Appendix \ref{ap_spectralresolution}, in this case the first-order moment
of the line is not modified by the spectrometer.

\subsection{Effect of a finite angular resolution}

Let us assume that we are observing with a telescope with a circularly symmetric
Gaussian beam with half-power beamwidth (HPBW) $\theta_b$.  
In the following we will use the beamwidth in units of $R_0$, i.e.\
$b\equiv D\,\theta_b/R_0$, where $D$ is the distance to the source.

The observed intensity as a function of projected distance will be the 2-D
convolution of $t(p)$, given by Eq.\ \ref{etp}, with the beam. 
Since the moments $\mu_0$ and $\mu'_1$ (but not $\mu_1$) depend linearly on the
line profile $t(p)$ (Eq.\ \ref{edefmu01}),  the convolution (a linear operator)
can be performed onto $\mu_0$ and $\mu'_1$, and after that, obtain the
normalized first-order moment $\mu_1(p;b)=\mu_0/\mu'_1$.

\begin{figure}[htb]
\centering
\resizebox{\hsize}{!}{\includegraphics{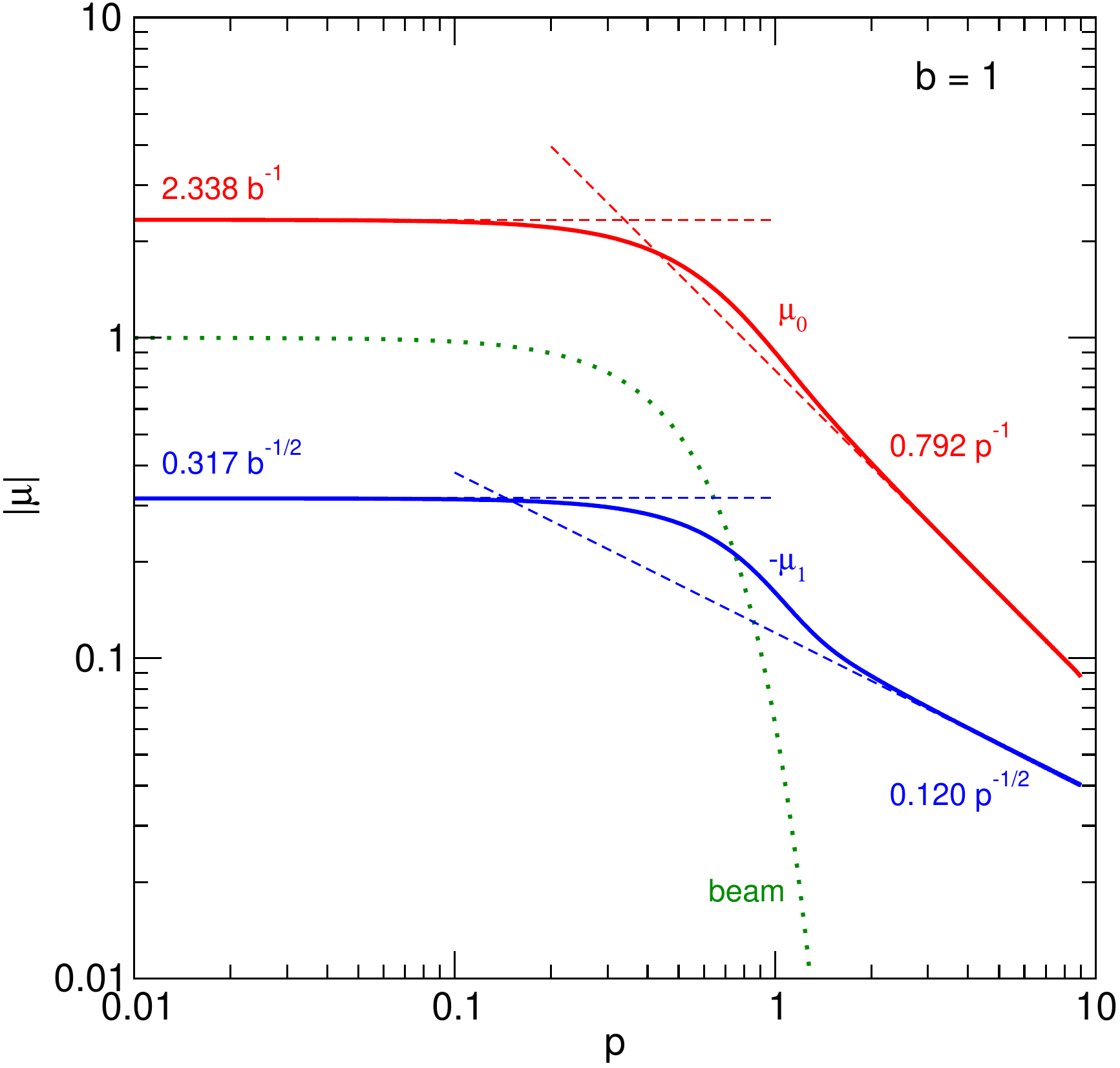}}
\caption{\label{fmu01_beta}
Log-log plots of the moments $\mu_0$ and $-\mu_1$ (solid lines) for a Gaussian
beam with HPBW $b=1$ (dotted line), as a function of the projected distance $p$,
and their asymptotic values (dashed lines) for $p\ll b$ and $p\gg b$. }
\end{figure}

The 2-D convolution of the power-law functions $\mu_0$ and $\mu'_1$ with a
Gaussian beam has to be done numerically. However, it has to be done only once,
because the result can be scaled for any value of $b$. For instance, if the
numerical calculation has been done for $b=1$ and $\mu_1(p;1)$ is obtained, then
for any value of $b$ we have
\begin{equation}
\mu_1(p;b)= b^{-1/2} \mu_1(p/b;1).
\end{equation}
The function $-\mu_1(p;1)$ is shown as a blue solid line in the log-log plot
of Fig.\ \ref{fmu01_beta}. 

However, since both $\mu_0$ and $\mu'_1$ are power laws of $p$,  
as shown in Appendix \ref{ap_convolution}, 
the value at the origin, $p=0$, and the asymptotic expression for large
projected distances, $p\gg b$, can be obtained analytically.
By application of Eqs.\ \ref{eqF0}, \ref{eqFs} to $\mu_0$ we obtain (see Fig.\
\ref{fmu01_beta})
\begin{eqnarray}
\mu_0(0;b) &=&      2.338\,b^{-1}, \nonumber\\
\mu_0(p;b) &\simeq& 0.792\,p^{-1} \quad (p\gg b),
\end{eqnarray}
with a characteristic size (the intersection of the two asymptotes) $p_0=
0.339\,b$.
For $\mu'_1$ we obtain
\begin{eqnarray}
{\mu'_1}(0;b) &=&      -0.742\,b^{-3/2}, \nonumber\\
{\mu'_1}(p;b) &\simeq& -0.095\,p^{-3/2} \quad (p\gg b),
\end{eqnarray}
with a characteristic size $p_1= 0.254\,b$.
Finally, for the first-order moment, $\mu_1=\mu'_1/\mu_0$, we derive
\begin{eqnarray}\label{eqmu1_0}
\mu_1(0;b) &=&      -0.317\,b^{-1/2}, \nonumber\\
\mu_1(p;b) &\simeq& -0.120\,p^{-1/2} \quad (p\gg b).
\end{eqnarray}

\section{First-order moment for a finite infall radius}

Up to now we have assumed that the infall velocity pattern in $r^{-1/2}$ extends
up to an infinite radius. A more realistic approach is to assume a finite
radius. For instance, in the inside-out collapse model \citep{Shu77}, 
the collapse propagates outwards from the center at the speed of sound. 
The radius of the expansion wave is usually called the infall radius $R_i$, and
let us call $r_i$ the infall radius in units of $R_0$, i.e.\ $r_i\equiv R_i/R_0$.
The envelope with a radius greater than the infall radius is approximately
static, while the material inside the infall radius is in free fall. Thus, let
us assume that the infall occurs only for radii $r<r_i$. The static material
will only contribute to the ambient-gas line-emission, centered on $v_z=0$, and
will not be taken into account.

\begin{figure}[htb]
\centering
\resizebox{\hsize}{!}{\includegraphics{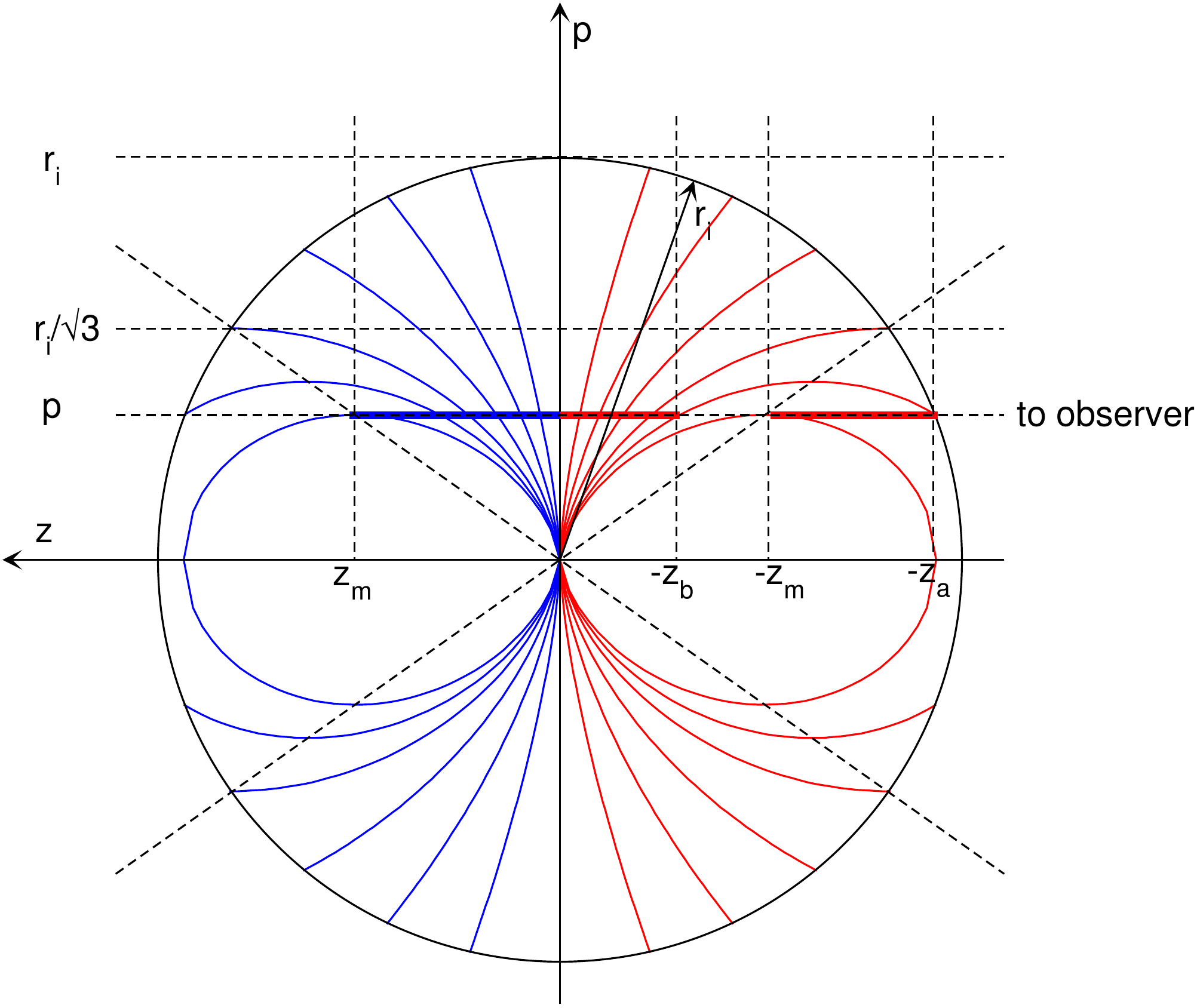}}
\caption{\label{feggs_ri}
For a given projected distance $p<r_i/\sqrt{3}$  the wing emission comes from
the thick part of the line of sight, for radii less than $r_i$. 
The emission from the thin part of the line of sight is hidden to the observer.
For the blue wing, the observer can observe the emission coming  from $z=0$
(with $v_z=0$) to $z=z_m$ (with $v_z=-v_m$). 
For the red wing, the observer can observe the emission coming from
$z=-z_a$ (with $r=r_i$, $v_z=v_a$) to  
$z=-z_m$ (with $v_z=+v_m$), 
and, since the corresponding part of the
equal-LOS-velocity surface facing the observer is missing (it would be outside
the infall radius), from 
$z=0$ (with $v_z=0$) to 
$z=-z_b$ (with $v_z=v_a$).
The velocity $v_a$ is the LOS velocity of the equal-LOS-velocity surface that
intersects the sphere $r=r_i$ at a projected distance $p$, i.e.\
$v_a=v_z(p, -z_a)=v_z(p, -z_b)$.
}
\end{figure}

\subsection{Line profile (finite infall radius)}

\begin{figure}[htb]
\centering
\resizebox{\hsize}{!}{\includegraphics{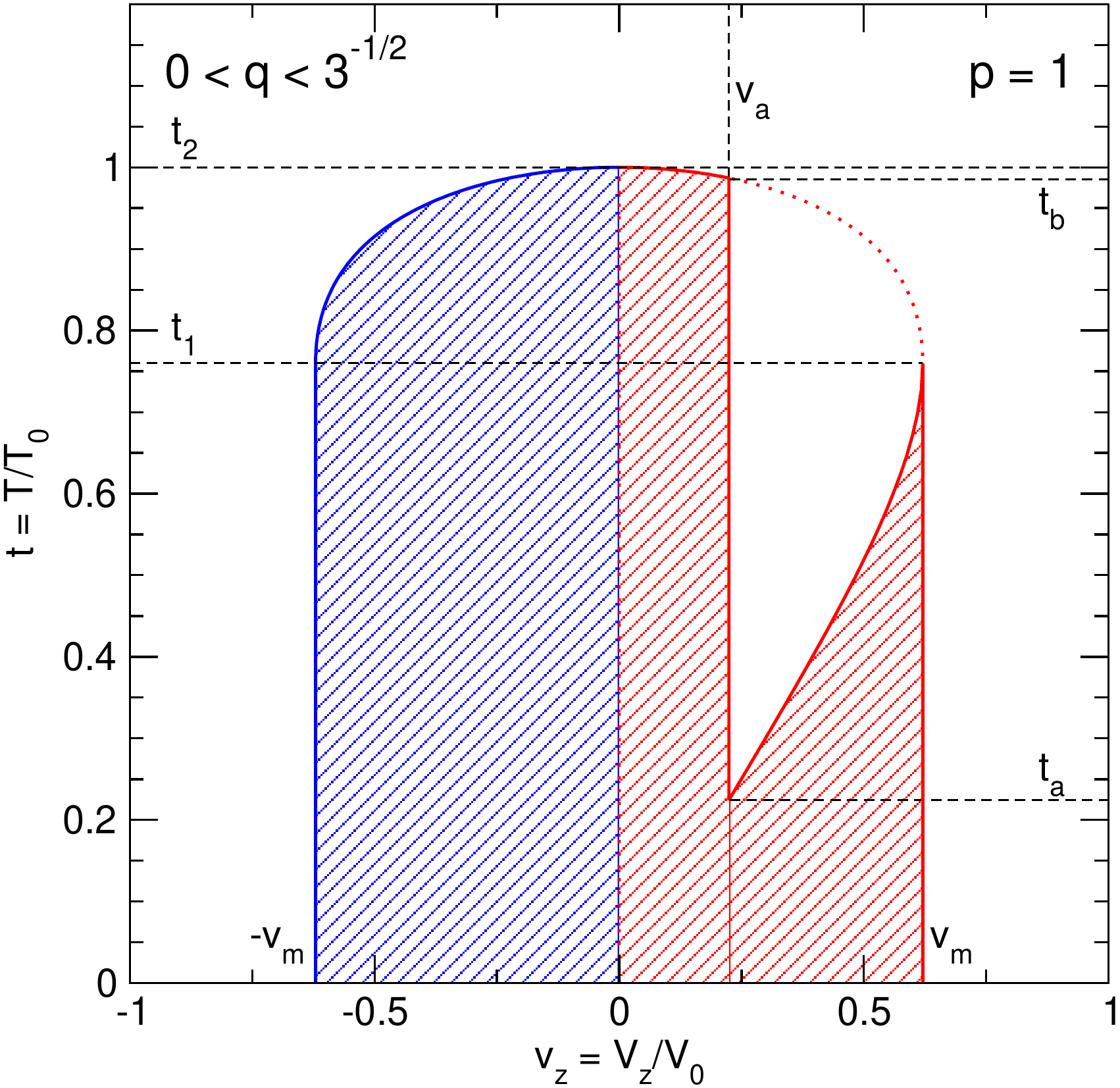}}
\caption{\label{flineprofile_ri}
Line profile for $0<q<1/\sqrt{3}$ and $p=1$. The blue wing is unaffected. For
the red wing the emission at velocities $v_z<v_a$ comes from the hot part of the
red equal-LOS-velocity surface. For $q\ge1/\sqrt{3}$ the blue and red wings
become symmetric.
}
\end{figure}

The effect of having a finite infall radius is that the equal-LOS-velocity
surfaces are truncated at a radius $r=r_i$. Thus, for the red-wing emission,  a
part of the equal-LOS-velocity surface near the center of the core will no
longer be hidden from the observer by the part facing the observer. A critical
value of $p$ is that for which the sphere of radius $r_i$ intersects the
equal-LOS-velocity surfaces at the points $(p=p_m, z=z_m)$, that is,
$r_i^2=p_m^2+z_m^2=3 p_m^2$, or $p_m=r_i/\sqrt{3}$.
As illustrated in Fig.\ \ref{feggs_ri}, for a given projected distance
$p < r_i/\sqrt{3}$, 
the blue-wing emission, like the infinite infall radius case, 
comes from material along the line of sight at radii
$p<r<(p^2+z_m^2)^{1/2}=\sqrt{3}p$. 
However, the red-wing emission, unlike the infinite infall radius case,  comes
from two regions:  material along the line of sight at radii $\sqrt{3}p<r<r_i$,
in the part of the equal-LOS-velocity surface facing the observer, and material
closer to the center, located at $-z_b<z<0$, which is no longer hidden by the
part of the equal-LOS-velocity surface facing the observer because this part of
the equal-LOS-velocity surface would be outside the infalling sphere of radius
$r_i$. The material in the part of the equal-LOS-velocity surface facing the
observer has LOS velocities $v_a<v_z<v_m$,  where $v_a$ is the velocity of the
material where the line of sight intersects the sphere of radius $r_i$.  The
material closer to the center has LOS velocities $0<v_r<v_a$  (see Fig.\
\ref{flineprofile_ri}).

Let us use the reduced coordinate $q\equiv p/r_i$. The corresponding critical
value of $q$ is $q=1/\!\sqrt{3}$ (see Fig.\ \ref{feggs_ri}).  For values
$q<1/\!\sqrt{3}$, only the red-wing emission is affected.  For
$q\ge1/\!\sqrt{3}$  the line becomes symmetric because none of the red wing
emission at any LOS velocity is hidden by the part of the equal-LOS-velocity
surface facing the observer, and the moment $\mu_1$ becomes zero (see Figs.\ 
\ref{feggs_ri} and \ref{flineprofile_ri}). For $q\ge1$ all the wing emission
disappears ($\mu_0=\mu_1=0$).

\subsection{First-order moment (finite infall radius)}

\begin{figure}[htb]
\centering
\resizebox{\hsize}{!}{\includegraphics{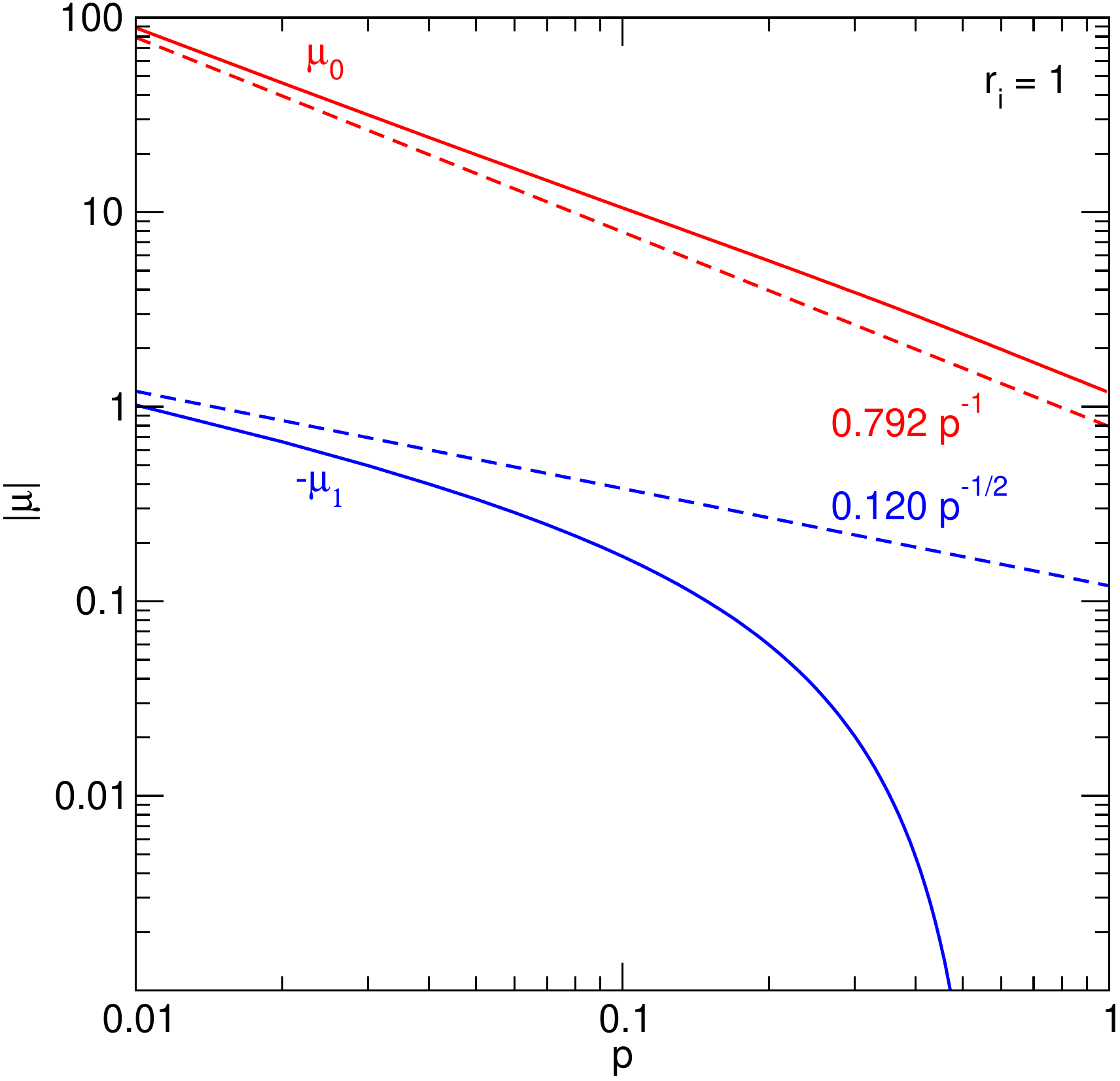}}
\caption{\label{fmu01_ri}
Log-log plot of $\mu_0$ and $-\mu_1$ as a function of the projected  distance
$p$, for an infall radius $r_i=1$ (solid lines) and for an infinite infall
radius (dashed lines).
}
\end{figure}

The details of the derivation of an analytical expression for the  moments, for
a finite infall radius are given in Appendix \ref{ap_finite}. The moments
$\mu_0$, $\mu'_1$, and $\mu_1$ obtained are no longer power laws of the
projected distance $p$, although the dependence on $r_i$ can be separated from
the explicit dependence on $p$ using the parameter $q= p/r_i$. In this way,
the resulting expressions are similar to those obtained for the case of an
infinite infall radius,  
\begin{eqnarray}\label{emu1_ri}
&&\left.\begin{array}{l}
\mu_0(p)  = H_0(q)\,p^{-1} \\
\mu'_1(p) = H_1(q)\,p^{-3/2} \\
\mu_1(p)  = [H_1(q)/H_0(q)]\,p^{-1/2}
\end{array}\right\}  \quad (0\leq q<1/\sqrt{3}), \nonumber\\
&&\left.\begin{array}{l} 
\mu_0(p) = 2B_0\,p^{-1}  \\
\mu'_1(p) = \mu_1(p)=0
\end{array}\right\}  \quad (1/\sqrt{3}\leq q<1), \\
&&\begin{array}{l}
\mu_0(p) = \mu'_1(p) = \mu_1(p) =0 
\end{array} \quad (q\ge1), \nonumber
\end{eqnarray}
with $H_0$ and $H_1$ given by (see Fig. \ref{fh01_ri})
\begin{eqnarray}
H_0(q) &=& 2B_0+[q(1-q^2)]^{1/2}({q'}^{1/2}-q^{1/2})-G_0(q)+G_0(q'),\nonumber \\
H_1(q) &=& \frac{q}{2}(1-q^2)({q'}^{1/2}-q^{1/2})-G_1(q)+G_1(q'),
\end{eqnarray}
where 
$q'$ is an auxiliary parameter depending on $q$ only 
(see Fig.\ \ref{fqprime}),
\begin{equation}
q'= \frac{1-q^2}{\left[1-\dfrac{q^2}{2}+\dfrac{q}{2}(4-3q^2)^{1/2}\right]^{1/2}},
\end{equation}
and
\begin{eqnarray}
G_0(y) &=& \frac{\pi}{8}-\frac{y}{4}\sqrt{1-y^2}-\frac{1}{4}\arcsin y, \nonumber\\
G_1(y) &=& \frac{2}{21} - 
           \frac{y^{3/2}}{2}\left(\frac{1}{3}-\frac{y^2}{7}\right),  \\
B_0    &=& \frac{\sqrt{2}}{4}+\frac{\pi}{8}-\frac{1}{4}\arcsin\frac{1}{\sqrt{3}}
            = 0.592. \nonumber
\end{eqnarray}
An example of the moments obtained can be seen in Fig.\ \ref{fmu01_ri}.

The limiting values of the moments for $r_i\to\infty$, corresponding to $q=0$,
$q'=1$,  coincide with the results derived for an infinite infall radius (Eqs.\ 
\ref{eqresult1} and \ref{eqresult2}),
\begin{equation}
\begin{array}{ll}
H_0(0)=H_0=2B_0-\pi/8,              & \mu_0= 0.792\,p^{-1},\\
H_1(0)=H_1=-\dfrac{2}{21}= -0.0952, & \mu_1= -0.120\,p^{-1/2},
\end{array}
\end{equation}
while the limiting values for $p=r_i/\sqrt{3}$, corresponding to  
$q=q'=1/\sqrt{3}$, are 
\begin{equation}
\begin{array}{ll}
H_0(1/\sqrt{3})=2B_0, & \mu_0= 1.185\,p^{-1},\\
H_1(1/\sqrt{3})=0,    & \mu_1= 0.
\end{array}
\end{equation}

\subsection{Finite angular resolution (finite infall radius)}

\begin{figure}[htb]
\centering
\resizebox{\hsize}{!}{\includegraphics{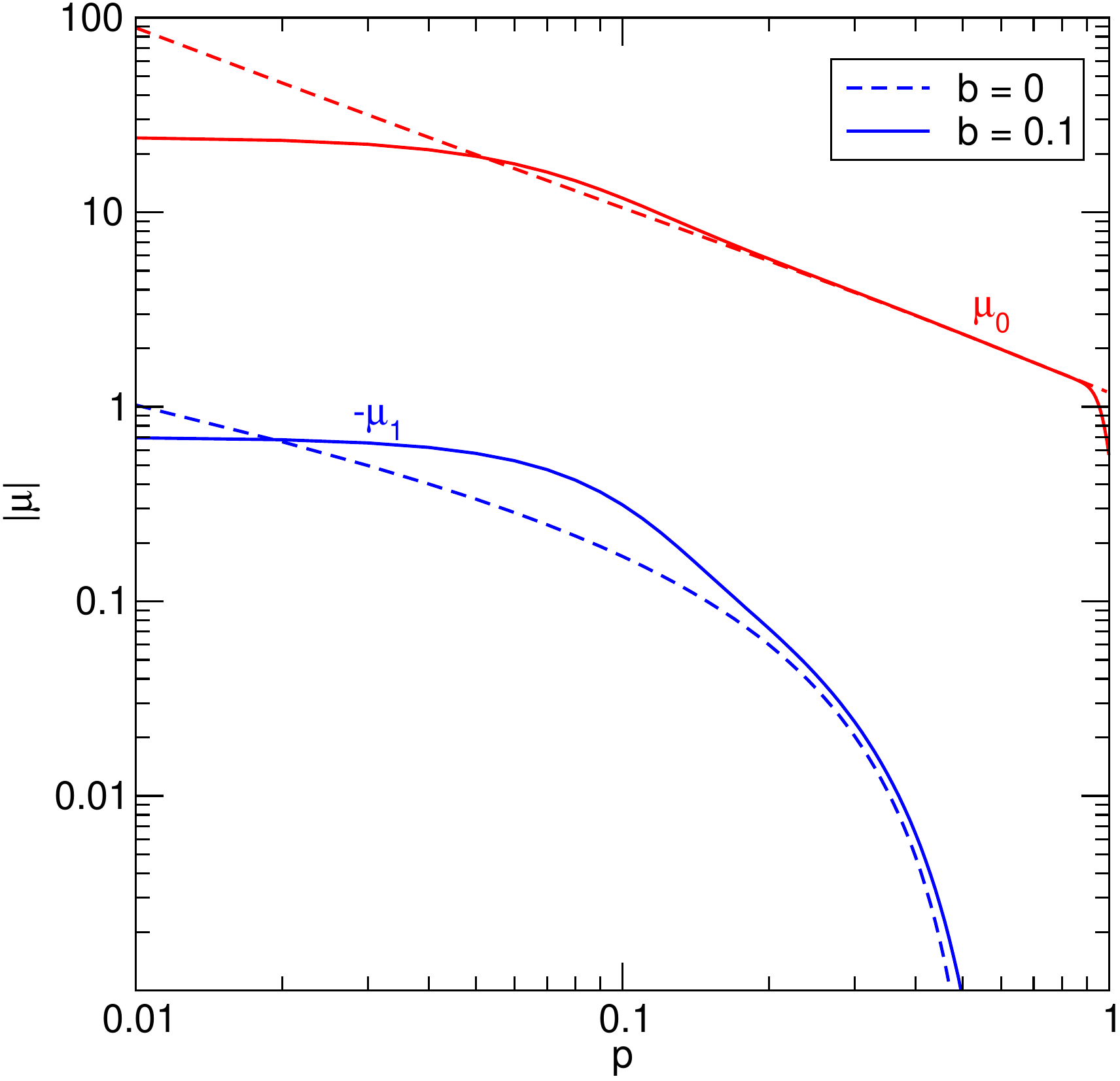}}
\caption{\label{fmu01_beta_ri} Log-log plot of $\mu_0$ and $-\mu_1$ as a
function of the projected distance,  for an infall radius $r_i=1$ and
beamwidth  $b = 0.1$ in reduced units (solid lines)  and  $b = 0$ (dashed
lines). }
\end{figure}

In this case, $\mu_0$ and $\mu'_1$ are no longer power laws of $p$, but have a
characteristic scale size given by $r_i$. Thus, unlike the infinite infall
radius case, the beam convolution has to be performed for every value of $r_i$,
and the result will depend on both $r_i$ and $b$. For $r_i$ very large, the
results for an infinite infall radius of the last section are reproduced.

\section{Practical units}

In the previous analysis 
all lengths, $z$, $p$, $r_i$, $b$, are measured in units of the reference radius
$R_0$; 
velocities in units of $V_0$, the infall velocity at the reference radius
$R_0$; 
and temperatures in units of $T_0$, the temperature at the reference radius
$R_0$. 
The projected distance $p$ in practical units, i.e.\ $\theta$ in arcsec, can be
obtained from $p$ in units of $R_0$ through\footnote{$1 \mathrm{~kau} = 1000
\mathrm{~au} = 0.0048$ pc, is a convenient unit when dealing with typical
circumstellar sizes. It is the physical size corresponding to an angular size of
$1''$ at a distance of 1 kpc.}
\begin{equation}\label{eqpd}
\left[\frac{\theta}{\mathrm{arcsec}}\right]= p 
\left[\frac{R_0}{\mathrm{kau}}\right]
\left[\frac{D}{\mathrm{kpc}}\right]^{-1},
\end{equation}
where $D$ is the distance to the source.
The value of $R_0$ is arbitrary, but the values of $R_0$ and $V_0$ are related
to the central mass of the protostar onto which the accretion is taking place,
$M_\ast$,
\begin{equation}\label{eqMast}
\frac{G\,M_\ast}{R_0}= \frac{1}{2} V_0^2,
\end{equation}
or, in practical units,
\begin{equation}\label{eqr0}
\left[\frac{V_0}{\mathrm{km~s}^{-1}}\right]=1.331
\left[\frac{M_\ast}{M_\sun}\right]^{1/2}
\left[\frac{R_0}{\mathrm{kau}}\right]^{-1/2}.
\end{equation}

\subsection{Infinite infall radius}

\begin{figure}[htb]
\centering
\resizebox{\hsize}{!}{\includegraphics{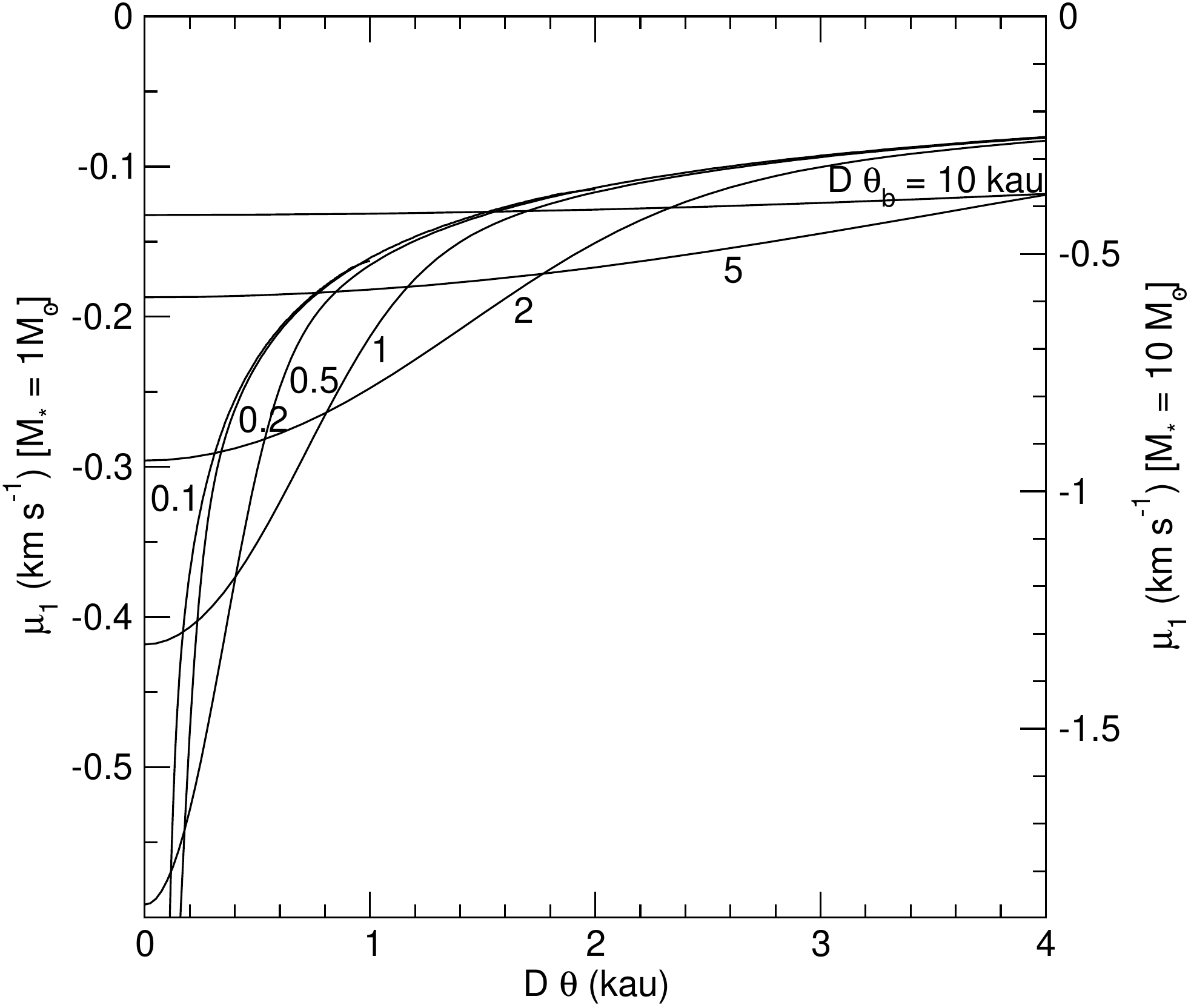}}
\caption{\label{fmu1_beta}
First-order moment $\mu_1$ for an infinite infall radius, 
as a function of the projected distance, $D\,\theta$ (in kau), 
for beamwidths  $D\,\theta_b = 0.1$, 0.2, 0.5, 1, 2, 5, and 10~kau, 
where $D$ is the distance to the source. 
The vertical axis scales as ${M_\ast}^{1/2}$.
The left axis is $\mu_1$ for a central mass $M_\ast=1$~\Msol,
while the right axis is for $M_\ast=10$~\Msol.
}
\end{figure}

\begin{figure}[htb]
\centering
\resizebox{\hsize}{!}{\includegraphics{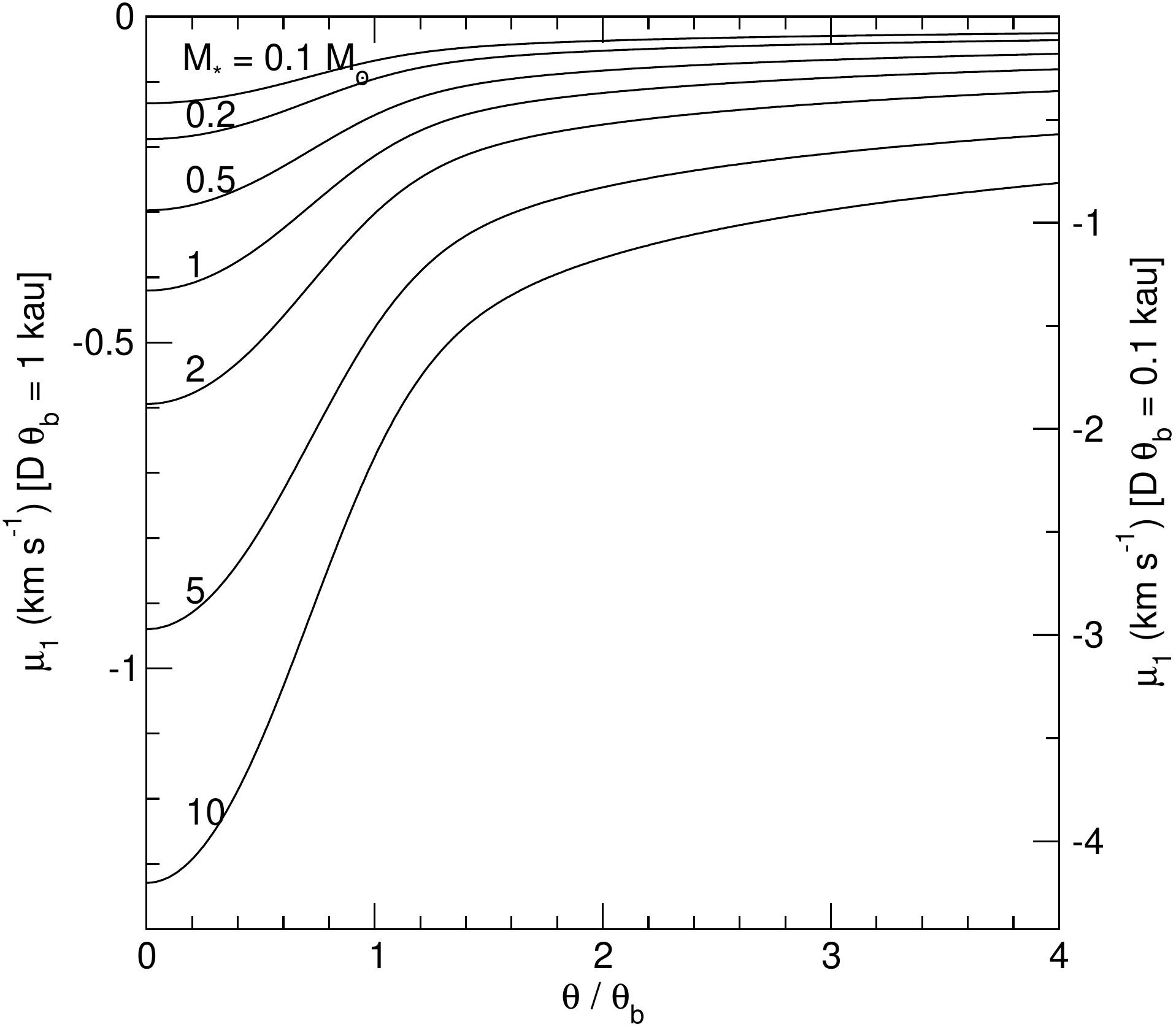}}
\caption{\label{fmu1_mass}
First-order moment  $\mu_1$ for an infinite infall radius, 
as a function of the projected distance in units of the beamwidth,
$\theta/\theta_b$ (adimensional), for central masses 
$M_\ast= 0.1$, 0.2, 0.5, 1, 2, 5, and $10$~\Msol. 
The vertical axis scales as $(D\,\theta_b)^{1/2}$, where $D$ is the distance to 
the source.
The left axis is $\mu_1$ for $D\,\theta_b=1$~kau,
while the right axis is for $D\,\theta_b=0.1$~kau.
}
\end{figure}

The first-order moment at the origin (Eq.\ \ref{eqmu1_0}) in practical units
becomes,
\begin{equation}
\left[\frac{\mu_1(0;\theta_b)}{\mathrm{km~s^{-1}}}\right] = -0.317
\left[\frac{V_0}{\mathrm{km~s^{-1}}}\right]
\left[\frac{R_0}{\mathrm{kau}}\right]^{1/2}
\left[\frac{D}{\mathrm{kpc}}\right]^{-1/2}
\left[\frac{\theta_b}{\mathrm{arcsec}}\right]^{-1/2},
\end{equation}
while, for $\theta\gg \theta_b$,
\begin{equation}
\left[\frac{\mu_1(\theta;\theta_b)}{\mathrm{km~s^{-1}}}\right] \simeq -0.120
\left[\frac{V_0}{\mathrm{km~s^{-1}}}\right]
\left[\frac{R_0}{\mathrm{kau}}\right]^{1/2}
\left[\frac{D}{\mathrm{kpc}}\right]^{-1/2}
\left[\frac{\theta}{\mathrm{arcsec}}\right]^{-1/2}.
\end{equation}
Taking into account Eq.\ \ref{eqMast}, we have
for $\theta=0$
\begin{equation}\label{eqmu10}
\left[\frac{\mu_1(0;\theta_b)}{\mathrm{km~s^{-1}}}\right] =
-0.423
\left[\frac{M_\ast}{M_\sun}\right]^{1/2}
\left[\frac{D}{\mathrm{kpc}}\right]^{-1/2} 
\left[\frac{\theta_b}{\mathrm{arcsec}}\right]^{-1/2}, 
\end{equation}
and for $\theta\gg \theta_b$,
\begin{equation}
\left[\frac{\mu_1(\theta;\theta_b)}{\mathrm{km~s^{-1}}}\right] \simeq 
-0.160
\left[\frac{M_\ast}{M_\sun}\right]^{1/2}
\left[\frac{D}{\mathrm{kpc}}\right]^{-1/2} 
\left[\frac{\theta}{\mathrm{arcsec}}\right]^{-1/2}. 
\end{equation}
Examples of $\mu_1(\theta;\theta_b)$ for some values of $\theta_b$ and $M_\ast$
are shown in Figs.\ \ref{fmu1_beta} and \ref{fmu1_mass}.
A value of the central mass can be derived directly from the value of the 
first-order moment at the origin,
\begin{equation}\label{eqmumass}
\left[\frac{M_\ast}{M_\sun}\right]=
5.6
\left[\frac{-\mu_1(0;\theta_b)}{\mathrm{km~s^{-1}}}\right]^2
\left[\frac{D}{\mathrm{kpc}}\right] 
\left[\frac{\theta_b}{\mathrm{arcsec}}\right]. 
\end{equation}

It may seem surprising that the value of the first-order moment at the origin (Eq.\
\ref{eqmu10}) does not depend on the temperature $T_0$. This is because $\mu_1$
is the normalized moment $\mu'_1/\mu_0$, and the dependence on $T_0$ of both 
$\mu'_1$ and $\mu_0$ cancels. However, $\mu_1$ does depend on the temperature
gradient, i.e.\ the power-law index of temperature law ($\beta$ in Eq.\
\ref{eqmu1alpha}). For instance, for $\beta=0$ (no temperature gradient),
$\mu_1$ is zero.

\subsection{Finite infall radius}

\begin{figure}[htb]
\centering
\resizebox{\hsize}{!}{\includegraphics{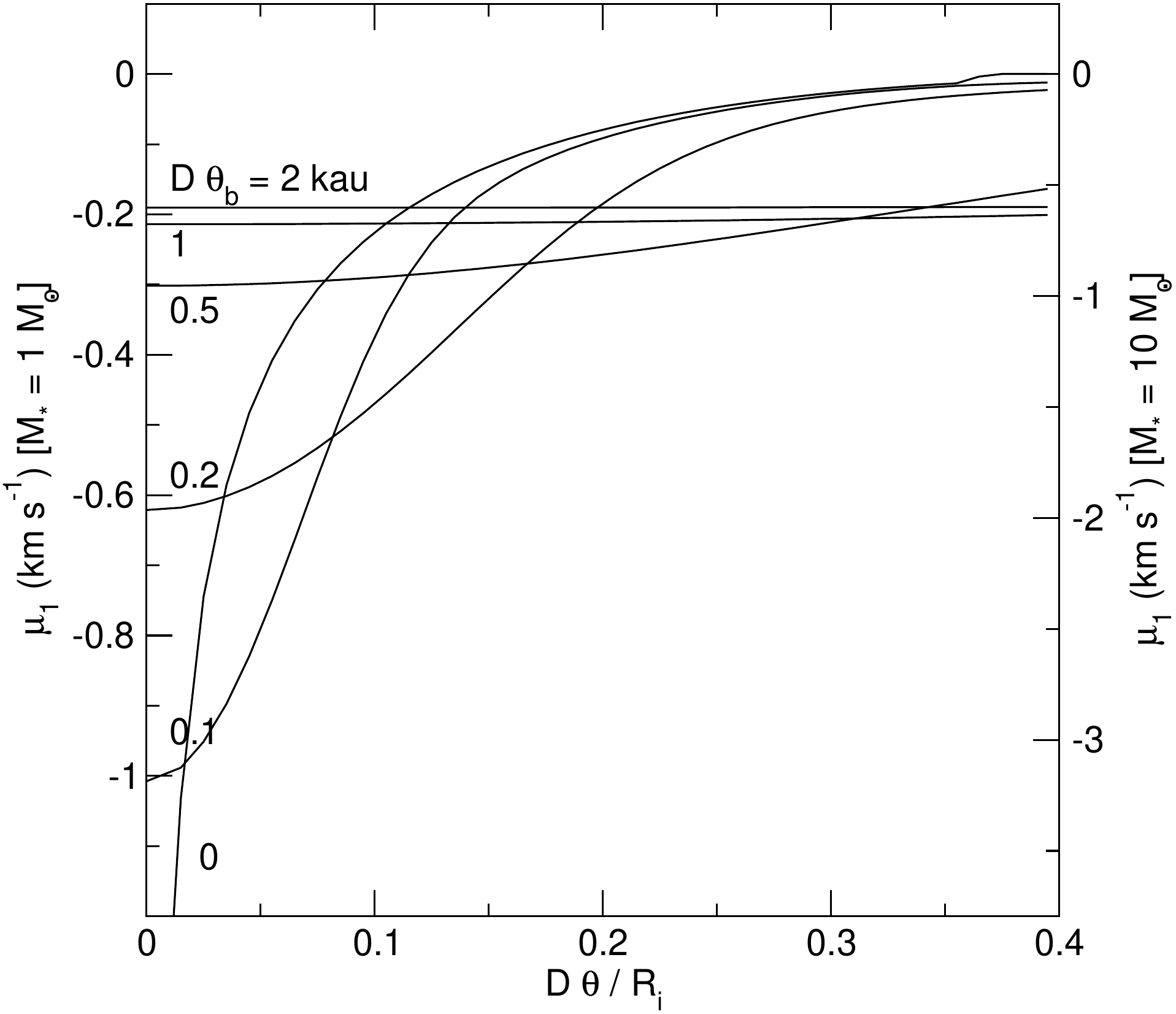}}
\caption{\label{fmu1_beta_ri}
First-order moment $\mu_1$, as a function of projected distance in units of the
infall radius, $D\,\theta/R_i$ (adimensional), for beamwidths
$D\,\theta_b= 0$, 0.1, 0.2, 0.5, 1, and 2~kau,
where $D$ is the distance to the source. 
The vertical axis scales as ${M_\ast}^{1/2}$.
The left axis is $\mu_1$ for a central mass $M_\ast=1$~\Msol,
while the right axis is for $M_\ast=10$~\Msol.
}
\end{figure}

\begin{figure}[htb]
\centering
\resizebox{\hsize}{!}{\includegraphics{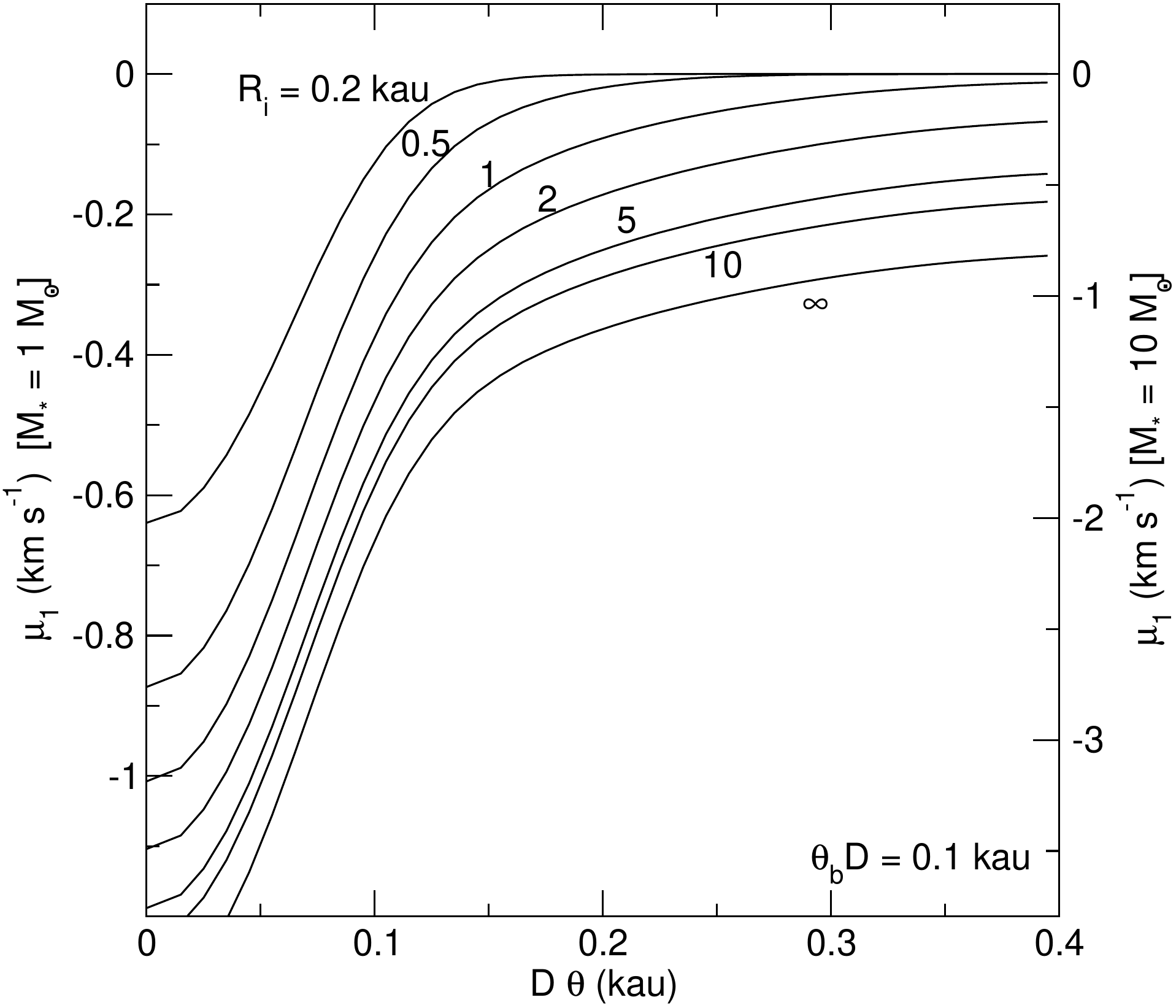}}
\caption{\label{fmu1_beta_ri2}
Same as Fig.\ \ref{fmu1_beta_ri}, for $D\,\theta_b=0.1$~kau, and  
$R_i= 0.2$, 0.5, 1, 2, 5, 10, and $\infty$ kau.
The horizontal axis is the projected distance, $D\,\theta$ (in kau).
Note that the line for $R_i=1$~kau in this figure coincides with the line for
$D\,\theta_b=0.1$~kau in Fig.\ \ref{fmu1_beta_ri}.
}
\end{figure}

In this case the 2-D convolution has to be computed for any value of the infall
radius. A possible strategy for computing the first-order moment is the following
\begin{enumerate}

\item Choose a value for $R_0$, for instance $R_0=1$~kau, so that the infall
radius in reduced coordinates is $r_i=
[R_i/\mathrm{kau}]$.

\item For a given value of $r_i$, and a range of values of $p$,
construct the functions $\mu_0=H_0(q)\,p^{-1}$ and $\mu'_1=H_1(q)\,p^{-3/2}$
(Eq.\ \ref{emu1_ri}) in reduced coordinates ($p$ and $r_i$ in units of $R_0$).

\item Transform the $p$
coordinate to practical units, 
$[\theta/\mathrm{arcsec}]=p\,[D/\mathrm{kpc}]^{-1}$ (Eq.\ \ref{eqpd}).
Compute the beam 2D-convolution (Eq.\ \ref{ebeam}) of $\mu_0$ and $\mu'_1$,
and obtain $\mu_1(\theta;\theta_b,R_i)=\mu'_1/\mu_0$.
See in Fig.\ \ref{fmu01_beta_ri} an example of the resulting
$\mu_0$ and $\mu_1$ after beam convolution. 

\item   The resulting first-order moment is in units of $V_0$. To have it in km~s$^{-1}$,
scale $\mu_1$ by a factor $1.331\,[M_\ast/M_\sun]^{1/2}$,  where $M_\ast$ is the
central mass (Eq.\ \ref{eqr0}).

\end{enumerate}
Examples of $\mu_1(\theta;\theta_b,R_i)$ for some values of $\theta_b$, $R_i$,
and $M_\ast$ are shown in Figs.\ \ref{fmu1_beta_ri} and Fig.\
\ref{fmu1_beta_ri2}. 

\begin{figure}[htb]
\centering
\resizebox{\hsize}{!}{\includegraphics{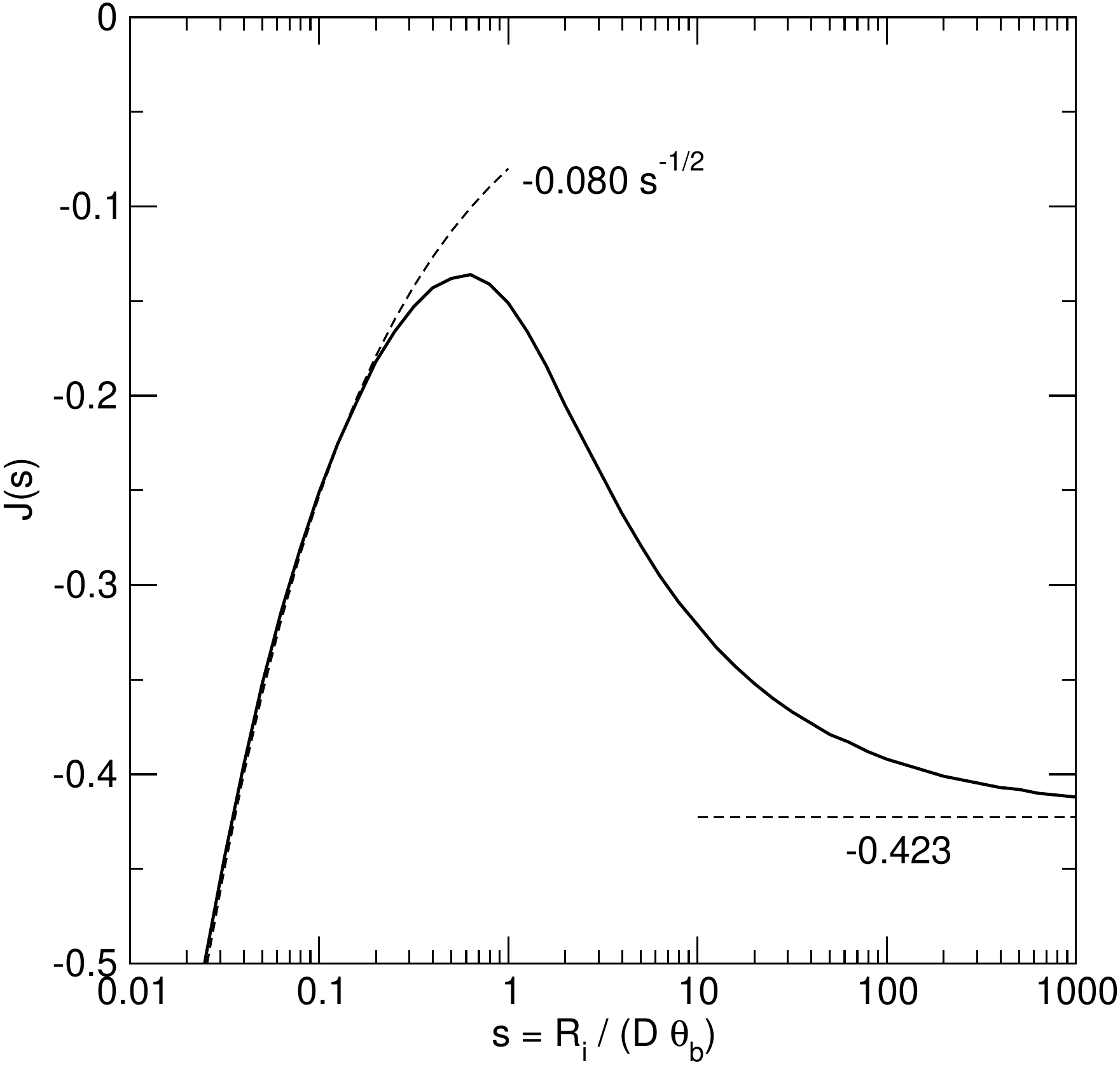}}
\caption{\label{figJ}
Function $J(s)$ in the expression of $\mu_1(0;\theta_b,R_i)$, showing its
asymptotic values for $s\ll1$ and $s\gg1$.}
\end{figure}

Let us 
see now what can be derived from the value of the first-order moment at the
origin.
The value of $\mu_1$ for $\theta=0$ (Eq.\ \ref{eqJ1J0}) in
practical units becomes
\begin{equation}
\left[\frac{\mu_1(0;\theta_b,R_i)}{\mathrm{km~s^{-1}}}\right] = 
\frac{J_1(s)}{J_0(s)}
\left[\frac{V_0}{\mathrm{km~s^{-1}}}\right]
\left[\frac{R_0}{\mathrm{kau}}\right]^{1/2}
\left[\frac{D}{\mathrm{kpc}}\right]^{-1/2} 
\left[\frac{\theta_b}{\mathrm{arcsec}}\right]^{-1/2}, 
\end{equation}
where $s=R_i/(D\theta_b)$, and $J_0$ and $J_1$ are given by Eq.\ \ref{eqJ}. 
Taking into account Eq.\ \ref{eqMast}, we have
\begin{equation}
\left[\frac{\mu_1(0;\theta_b,R_i)}{\mathrm{km~s^{-1}}}\right] = 
J(s)
\left[\frac{M_\ast}{M_\sun}\right]^{1/2}
\left[\frac{D}{\mathrm{kpc}}\right]^{-1/2} 
\left[\frac{\theta_b}{\mathrm{arcsec}}\right]^{-1/2}, 
\end{equation}
where $J(s)= 1.331\,J_1(s)/J_0(s)$. The function $J(s)$, calculated
numerically, is shown in Fig.\ \ref{figJ}. The asymptotic expression for large
infall radii,  $R_i\gg D\theta_b$,
\begin{equation}
\left[\frac{\mu_1(0;\theta_b,R_i\!\gg\!D\theta_b}{\mathrm{km~s^{-1}}}\right] \simeq
-0.423
\left[\frac{M_\ast}{M_\sun}\right]^{1/2}
\left[\frac{D}{\mathrm{kpc}}\right]^{-1/2} 
\left[\frac{\theta_b}{\mathrm{arcsec}}\right]^{-1/2}, 
\end{equation}
coincides, as expected, with the expression derived for an infinite infall
radius, Eq.\ \ref{eqmu10}. In the case of a poor angular resolution compared
with the infall radius ($\theta_b\gg R_i/D$), we obtain
\begin{equation}
\left[\frac{\mu_1(0;\theta_b\!\gg\!R_i/D,R_i)}{\mathrm{km~s^{-1}}}\right] \simeq
-0.080
\left[\frac{M_\ast}{M_\sun}\right]^{1/2}
\left[\frac{R_i}{\mathrm{kau}}\right]^{-1/2}. 
\end{equation}
The value at the origin of the first-order moment in the case of a finite infall
radius, $\mu_1(0;\theta_b,R_i)$, in contrast with the infinite infall radius
case. does not provide a unique value of the central mass, unless the infall
radius is known. There is a degeneracy between the infall radius and the central
mass: different pairs of values of the infall radius and central mass produce
the same value of the first-order moment at the origin. In order to
disentangle this degeneracy, it is necessary to fit not only the value at the
origin, but also the variation of the observed first-order moment as a
function of the projected distance.

\section{Application of the central-blue-spot infall hallmark to real data}

\subsection{G31.41+0.31 HMC}

\object{G31.41+0.31 HMC} 
(hereafter G31) is a hot molecular core whose distance was estimated to
be 7.9 kpc, but recent determinations (Reid et al.\ in preparation) give a
value of 3.7 kpc for its distance.
Infall motions in G31 have been reported by \citet{Gir09} from inverse
P-Cygni profiles, and by \citet{May14} from VLA observations of the ammonia
inversion transitions $(2,2)$, $(3,3)$, $(4,4)$, $(5,5)$, and $(6,6)$ showing
a central blue-spot in the first-order moment maps.

\begin{figure}[htb]
\centering
\resizebox{0.564\hsize}{!}{\includegraphics{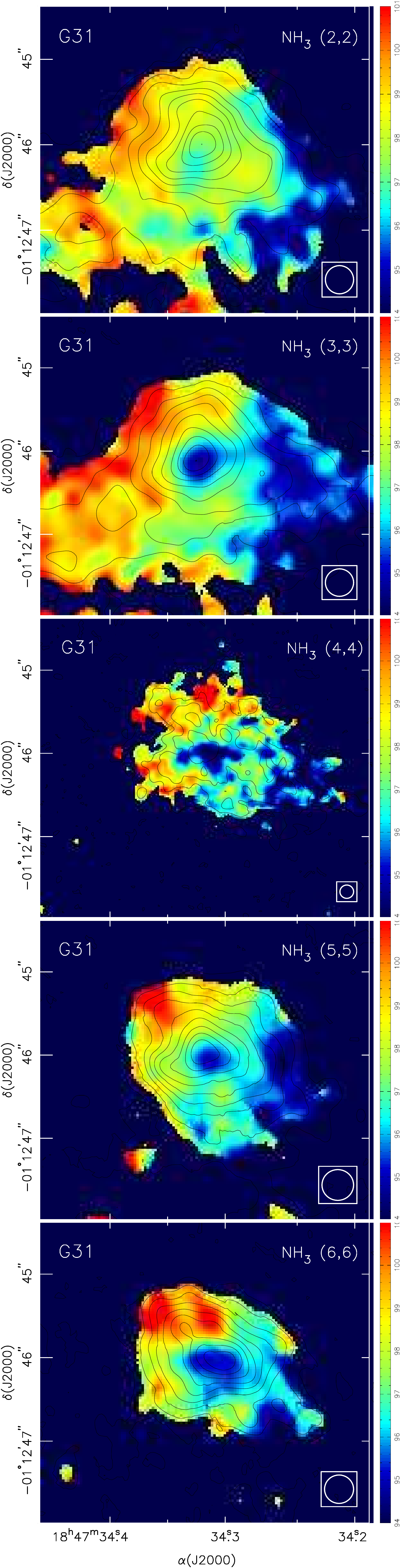}}
\caption{\label{g31_maps}
Maps of G31 first-order (color scale) and zeroth-order moment (contours) 
for the ammonia inversion transitions $(2,2)$ to
$(6,6)$ \citep{May14}. 
The color scale, the same for all panels, ranges from 94 to 101 \kms{}. 
The contours are in steps of 10\% of the peak value for all the maps, 
except for the $(4,4)$ transition, for which the steps are 20\%.
The beam is shown in the lower right corner of each panel.
}
\end{figure}

\begin{figure}[htb]
\centering
\resizebox{\hsize}{!}{\includegraphics{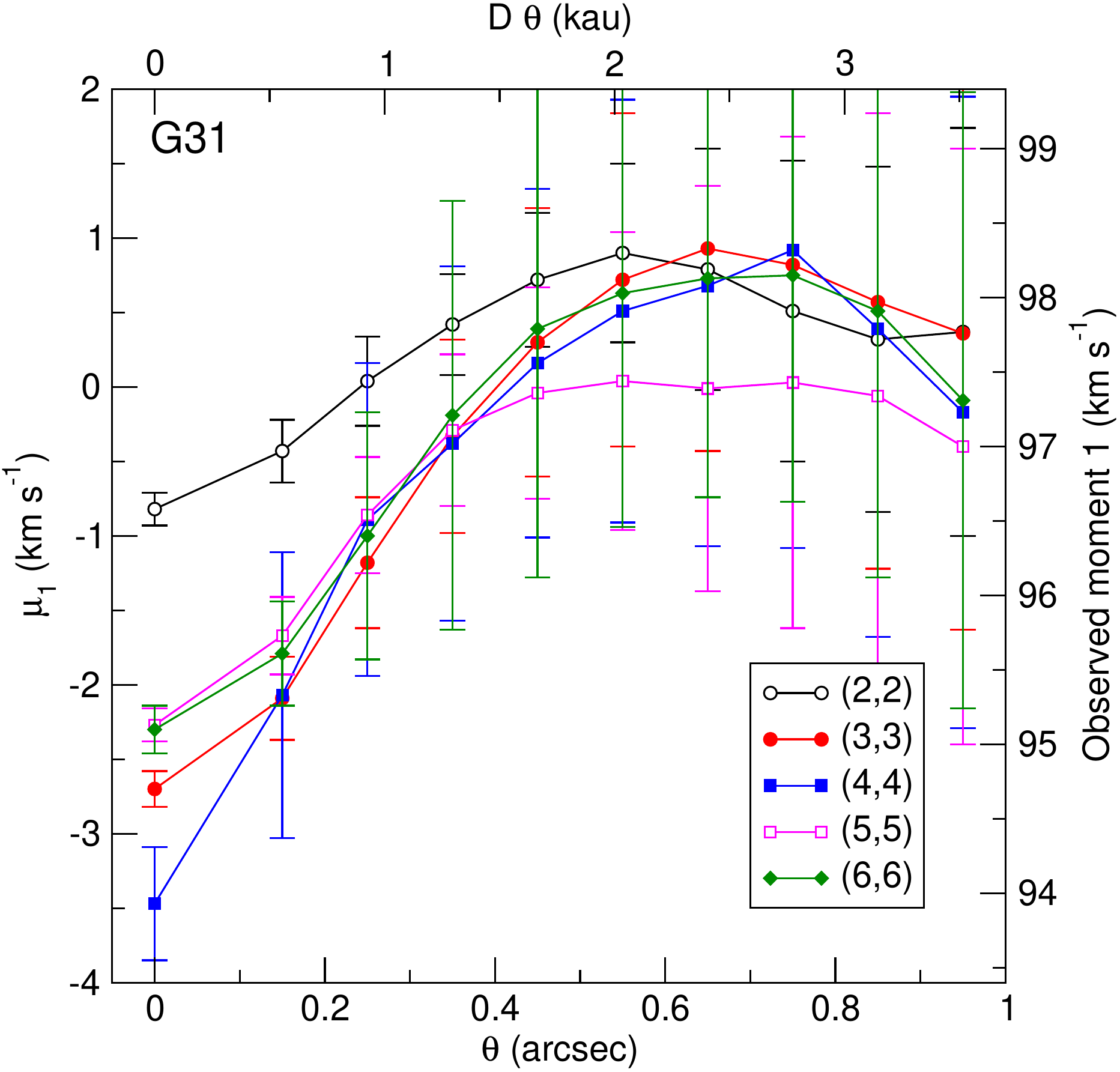}}
\caption{\label{g31_mom1}
G31 first-order moment for the ammonia
inversion transitions 
$(2,2)$ (black      line and  open circles),             
$(3,3)$ (red        line and  filled circles),           
$(4,4)$ (blue       line and  filled squares),           
$(5,5)$ (magenta    line and  and open squares), and      
$(6,6)$ (green      line and  filled diamonds)           
\citep{May14}, as a function of angular distance, measured for rings of
width $0\farcs1$ and average radius $\theta$.
The error bars are the rms of the velocity inside each ring. The right-hand
vertical axis shows the velocities obtained from the first-order 
moment maps, while the left-hand vertical axis shows the velocity with respect
to the systemic velocity of G31, taken as $V_\mathrm{sys}=97.4$ \kms.
}
\end{figure}

\begin{figure}[htb]
\centering
\resizebox{\hsize}{!}{\includegraphics{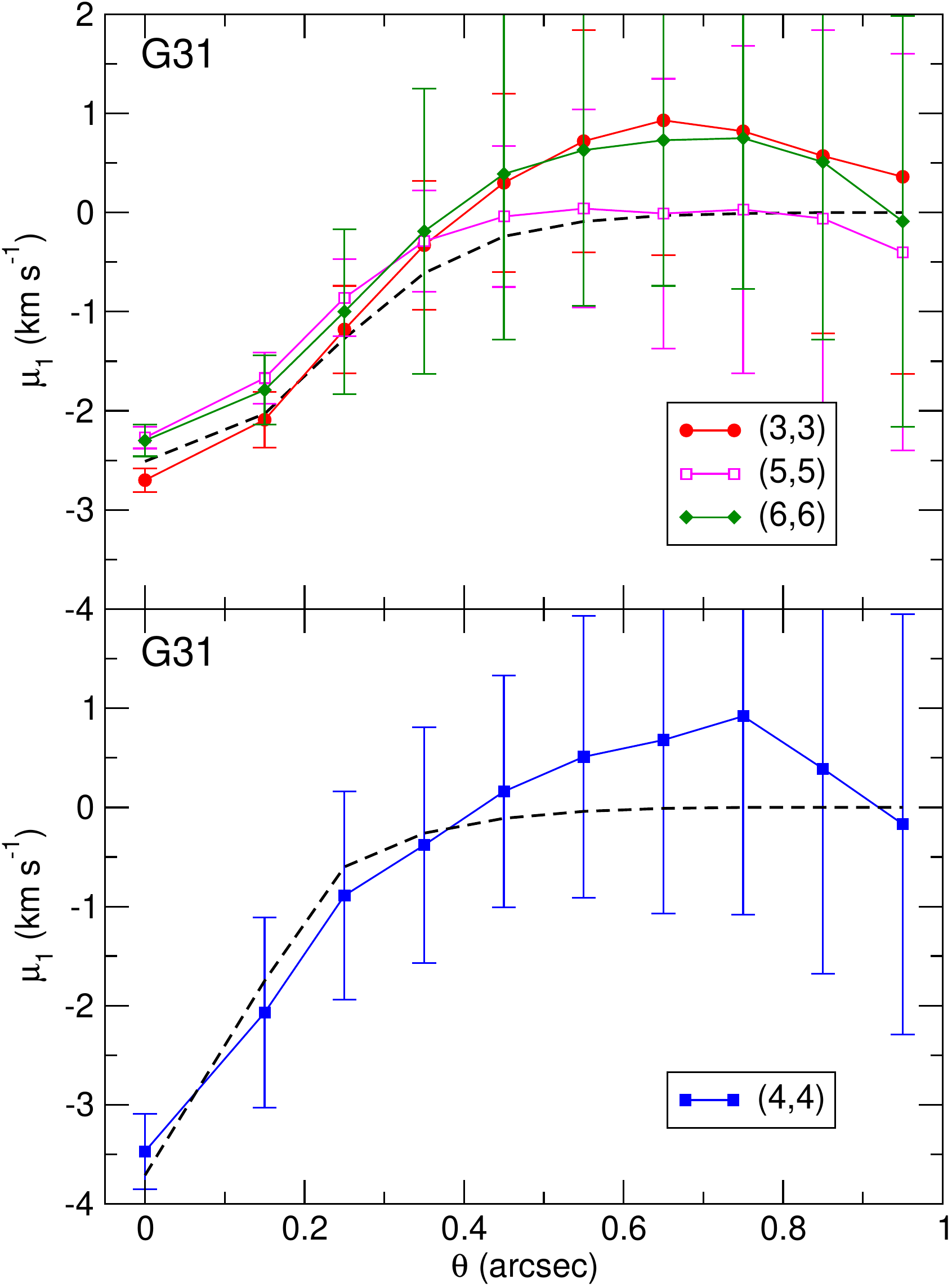}}
\caption{\label{g31_model}
Model (black dashed line) fitted to the G31 first-order moment, calculated for a
central mass of 122 \Msol, an infall radius of 5 kau, 
and a half-power beamwidth of $\sim0\farcs33$ for the
$(3,3)$, $(5,5)$, and $(6,6)$ transitions (top), and $0\farcs16$ for the $(4,4)$
transition (bottom).
}
\end{figure}

Here we are analyzing the first-order moment maps shown in Fig.\ 2 of
\citet{May14} (see Fig.\ \ref{g31_maps}). 
The half-power beamwidth of the observations were 
$0\farcs33$ for the $(2,2)$, $(3,3)$, and $(6,6)$ maps, 
$0\farcs16$ for the $(4,4)$ map,
and
$0\farcs37$ for the $(5,5)$ map. 
The value of the first-order moment as a function of the angular distance was
obtained for the five maps by averaging the first-order moment in concentric
rings of width $0\farcs1$ centered on the average position of the peak of the
blue spot, \RA{18}{47}{34}{32}, \DEC{-01}{12}{46}{1}.
In Fig.\ \ref{g31_mom1} we present the first-order moment profiles
obtained. We considered that the velocity far from the center was 97.4 \kms, the
value used by \citet{May14}, which is consistent with the values of the systemic
velocity quoted for G31, ranging from 96.26 \kms\ \citep{Bel05} to 98.8 \kms\
\citep{Ces94}.
The first-order moment of the different transitions show very similar profiles,
with a value of $\sim-3$ \kms\ for small angular distances, except for the
$(2,2)$ transition, which shows a shallower dip in velocity. This could be due
to a lower opacity of the $(2,2)$ line, and partial blending of the central line
with the inner satellite lines, and we will no longer consider this transition.
By application of Eq.\ \ref{eqmumass} we see that a value of $\sim-3$ \kms\ for
the first-order moment means roughly a central mass of the order of 50 \Msol.

In order to obtain a more accurate value of the central mass, 
the hallmark model was calculated for the beam size of each transition, and
fitted to the observed data for different values of 
the infall radius and  the central mass. 
For an infinite infall radius the best fit was found for a  central mass of 
44 \Msol, 
with a residual $\chi^2$ statistic for $\nu=39$ degrees of freedom (the total
number of rings of all transitions used in the fit, minus 1), 
$\chi^2= 36.6$, which gives a reduced $\chi_r=(\chi^2/\nu)^{1/2}= 0.97$.
For finite values of the infall radius we obtained higher values of the central
mass. 
For $R_i=20$ kau (corresponding to an angular radius of $5\farcs4$ at 3.7 kpc),
we obtained a better fit, with a central mass 
$M_\ast= 69$ \Msol\ ($\chi^2= 21.2$, $\chi_r= 0.74$),
while for $R_i=5$ kau ($1\farcs4$), a lower limit for the infall radius to be
consistent with the size of the NH$_3$ maps, the best fit was for a central mass
$M_\ast= 122$ \Msol\ ($\chi^2= 10.7$, $\chi_r= 0.52$)
(see Fig.\ \ref{g31_model}).
In conclusion, the central mass obtained is always greater than $\sim44$ \Msol,
and the best fit is obtained for values of the infall radius between 5 and 20
kau, with central masses between $\sim70$ and $\sim120$ \Msol. 

\citet{Oso09} model the central core of G31 and obtain, 
assuming a distance of 7.9 kpc, 
a central star with a mass of $\sim25$ \Msol,  
a mass accretion rate of $3\,10^{-3}$ \Msol\ yr$^{-1}$, and  
a total luminosity of $2\,10^5$ \Lsol.
The luminosity, scaled to the distance of 3.7 kpc adopted here, is $4.4\,10^4$
\Lsol. A single star with a mass equal to the central mass derived here would
have a luminosity two orders of magnitude higher. This apparent lack of
luminosity can be explained considering, as usually found in high-mass
star forming regions, that there is not a single high-mass star at the center of
G31, but a cluster of less massive stars. In the case of G31, recent
high-angular resolution ALMA continuum observations (Beltr\'an et al.\ in
preparation) reveal the presence of at least four cores at the center of G31.

If we assume that the stars of the cluster at the center of G31 have masses that
follow the Salpeter initial mass function \citep{Sal55} there would be a few
high-mass young stellar objects, probably associated with the four cores
detected, which could account for most of the total luminosity observed. A
higher number of low-mass young stellar objects yet undetected, with little
contribution to the overall luminosity, could total a mass of 70--120 \Msol\ for
the cluster.

\subsection{B335}

\object{B335} is an isolated Class 0 protostar with a bolometric luminosity of
$\sim 1$ \Lsol, at a distance of 105 pc \citep{Olo09}.  Several authors claim
the detection of infall \citep{Kur13,Eva15}. Here we are analyzing two
observations,  
one of H$^{13}$CO$^+$ ($J=1$--0) at 87 GHz, with a moderate angular resolution,
combining data from the 45 m Nobeyama telescope and the Nobeyama Millimetre
Array \citep{Kur13},  
and the other of $^{13}$CO ($J=2$--1) at 220 GHz, with very high angular
resolution, carried out with the Atacama Large Millimeter/Submillimeter Array
(ALMA) \citep{Yen15}.

\begin{figure}[htb]
\centering
\resizebox{\hsize}{!}{\includegraphics{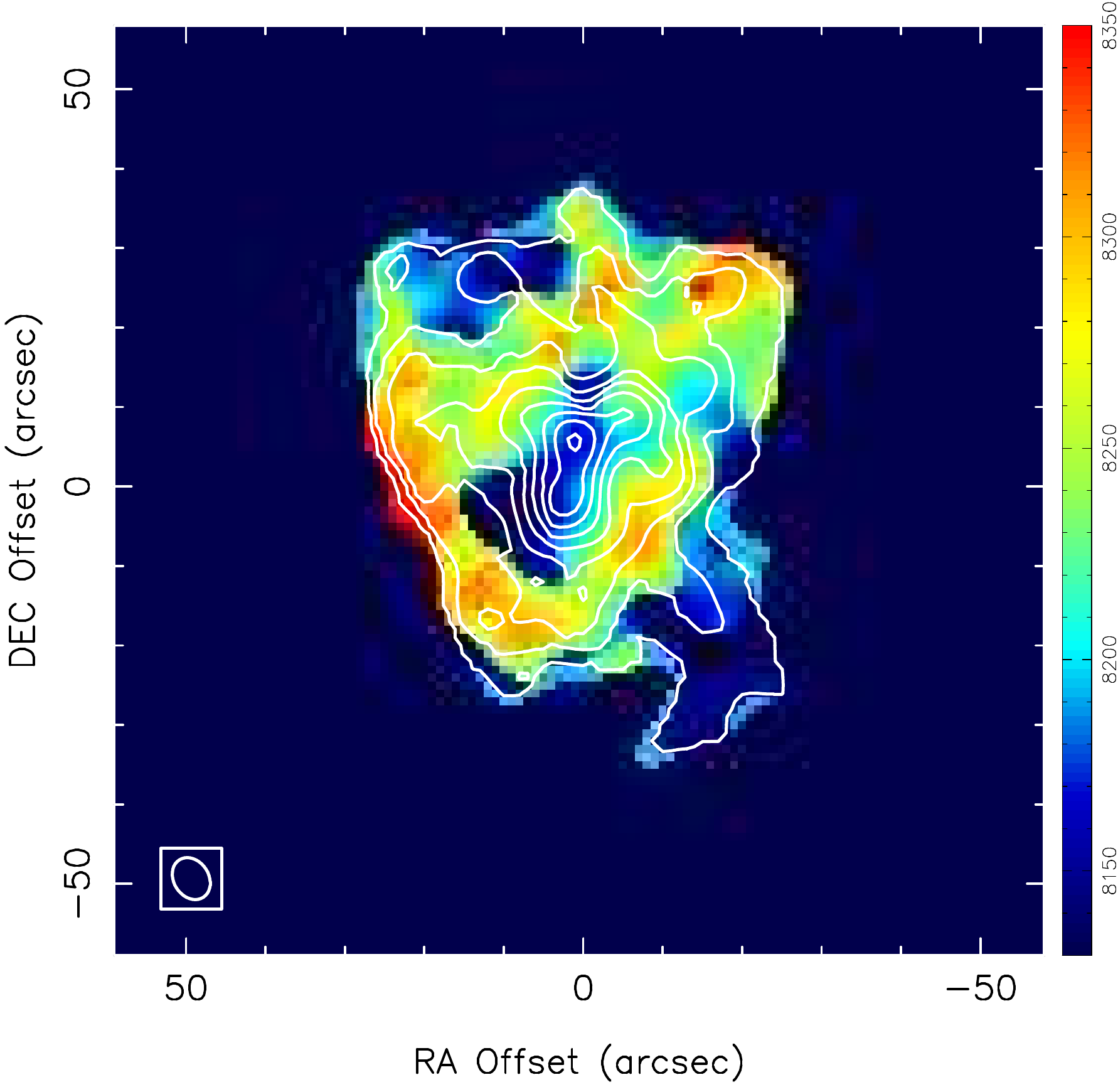}}
\caption{\label{b335_nobeyama_mom}
B335 zeroth-order (contours) and first-order moment (color
scale) of the H$^{13}$CO$^+$ ($J=1$--0) line, obtained from the channel maps of
\citet{Kur13}.
Contours are in steps of 511 mJy beam$^{-1}$ \kms.
The color scale at the right border is in m s$^{-1}$.
The $(0,0)$ position corresponds to \RA{19}{37}{00}{89}, \DEC{+07}{34}{09}{6}.
The synthesized beam is shown in the lower-left corner.
}
\end{figure}

\begin{figure}[htb]
\centering
\resizebox{0.5\hsize}{!}{\includegraphics{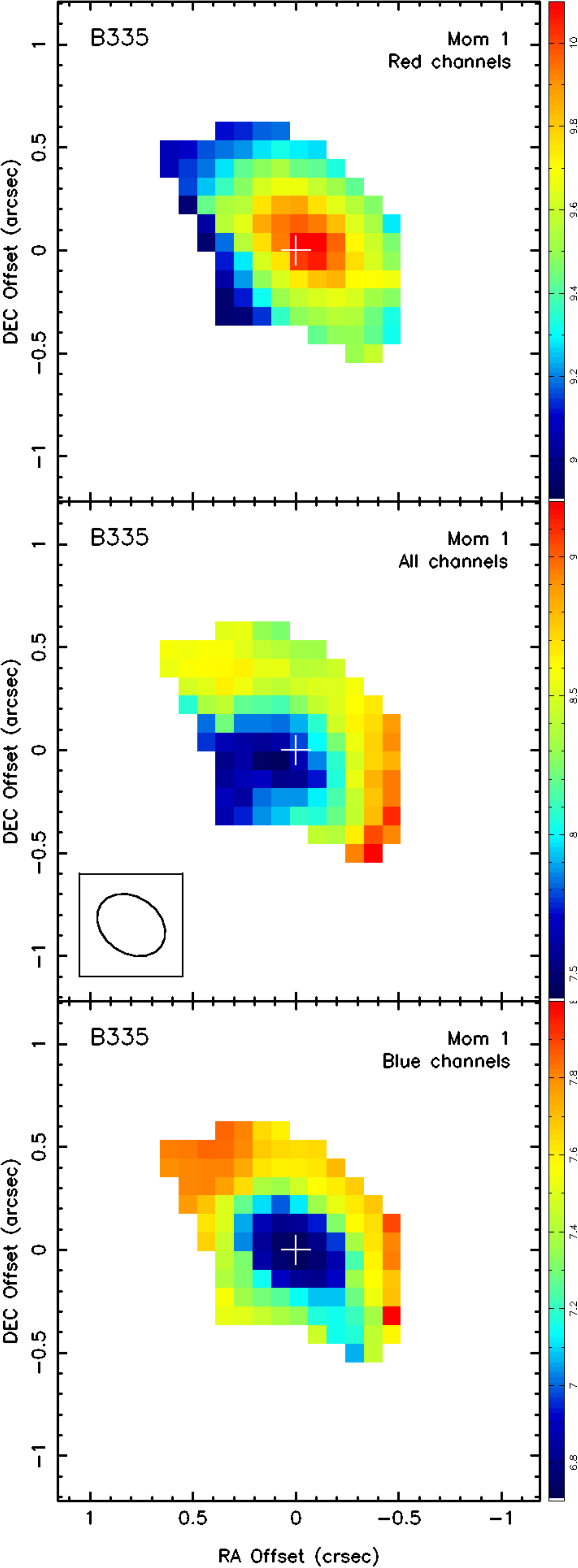}}
\caption{\label{fig_b335-alma}
B335 first-order moment of the $^{13}$CO ($J=2$--1) line observed with ALMA.
\textit{Top:} moment of the red channels, with velocities higher than the systemic
velocity, $V_\mathrm{sys}$ (color scale: 8.9--10.1 \kms). 
\textit{Middle:} moment of all the channels (color scale: 7.4--9.2 \kms).
\textit{Bottom:} moment of the blue channels, with velocities lower than
$V_\mathrm{sys}$ (color scale: 6.8--8.0 \kms).
The $(0,0)$ position, marked with a plus sign, 
is the same as in Fig.\ \ref{b335_nobeyama_mom}.
The synthesized beam is shown in the lower-left corner of the middle panel.
}
\end{figure}

\begin{figure}[htb]
\centering
\resizebox{0.7\hsize}{!}{\includegraphics{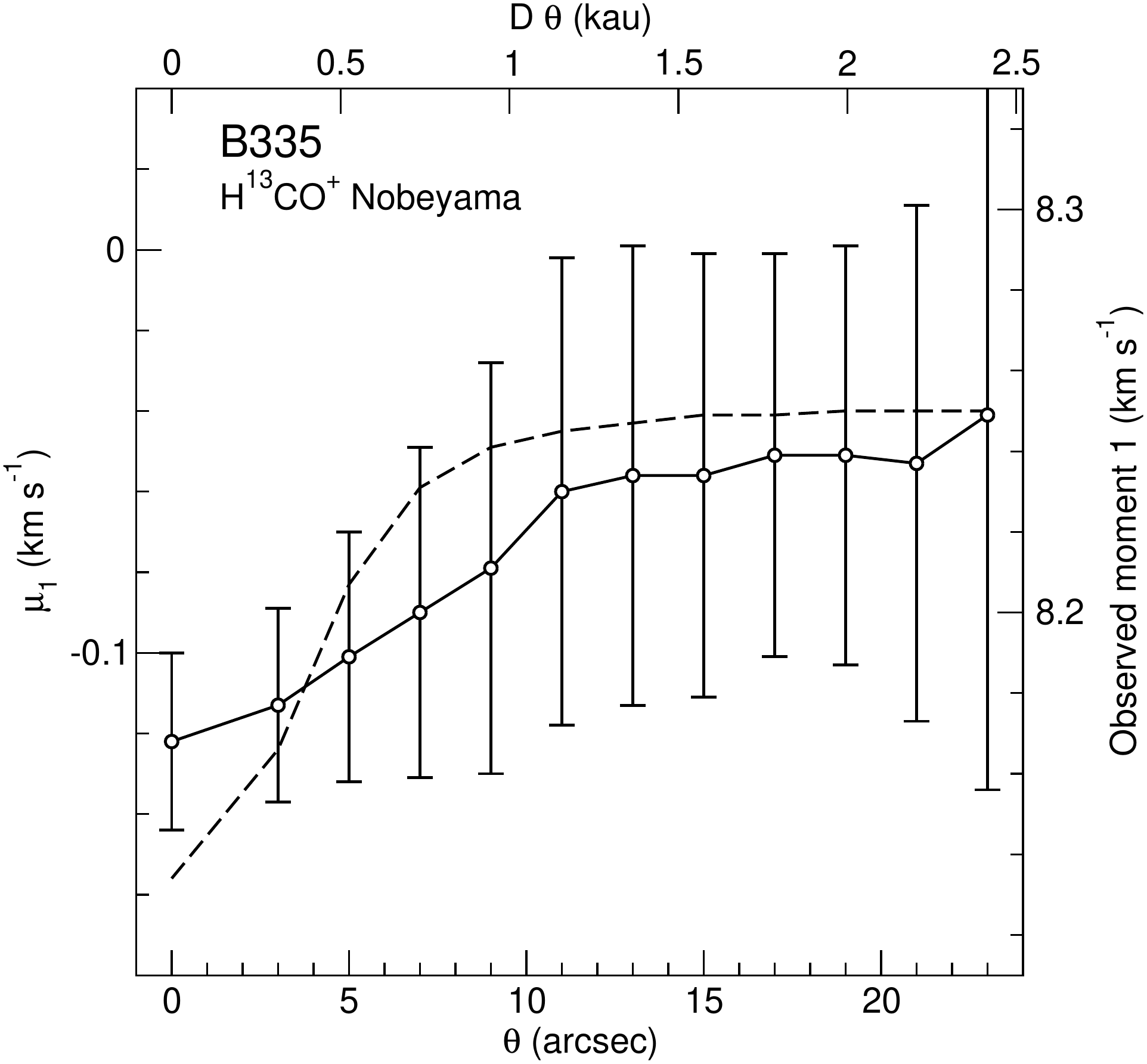}}
\resizebox{0.7\hsize}{!}{\includegraphics{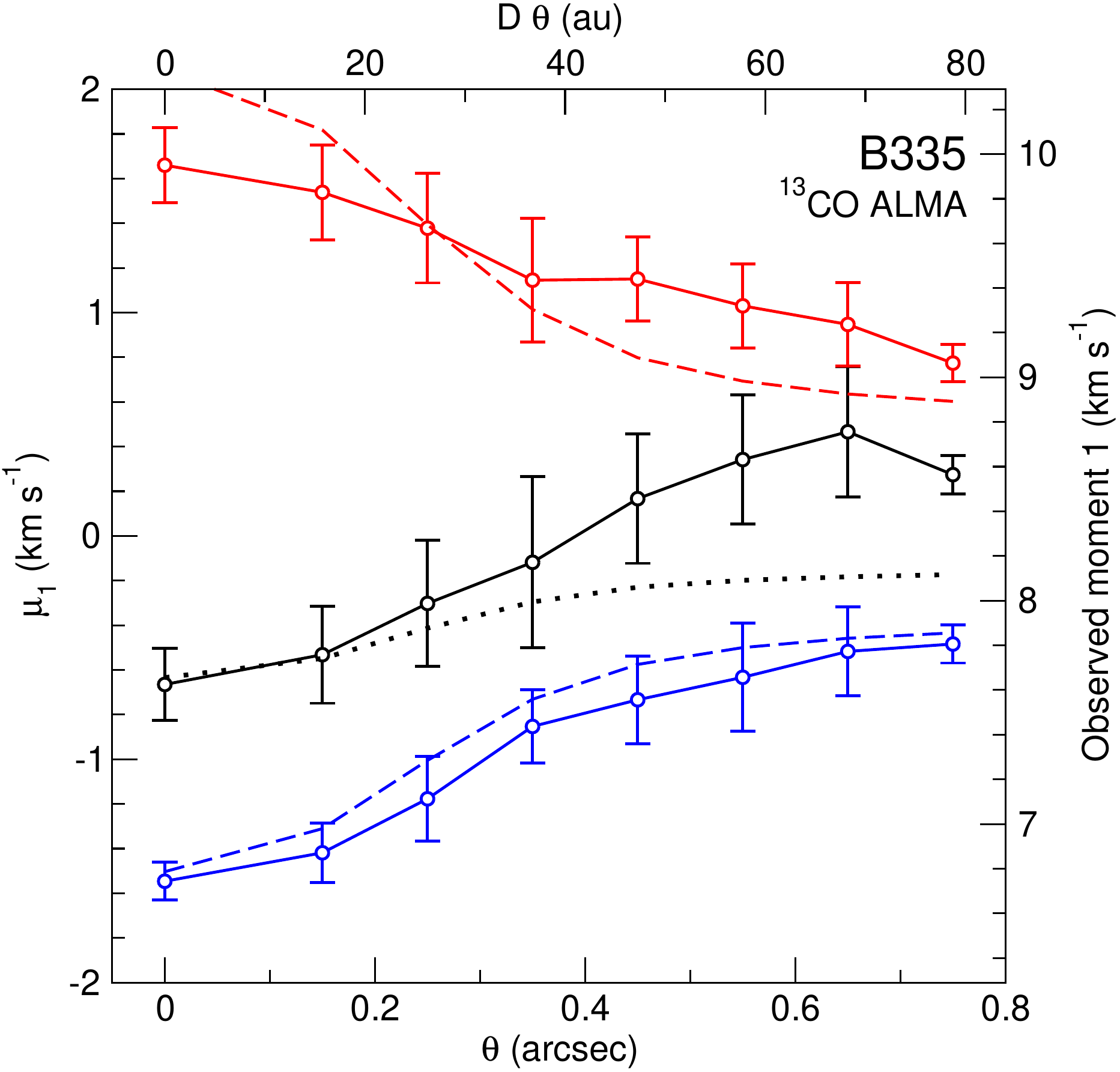}}
\caption{\label{b335_model}
Same as Fig.\ \ref{g31_mom1} for B335. 
\textit{Top:} H$^{13}$CO$^+$ ($J=1$--0) line observed at Nobeyama 
($2''$ ring width).
\textit{Bottom:} $^{13}$CO ($J=2$--1) line observed with ALMA 
($0\farcs1$ ring width;
$\mu_1^\mathrm{red}$, red line and symbols; 
$\mu_1$, black line and symbols; 
$\mu_1^\mathrm{blue}$, blue line and symbols) 
The dashed lines in the top and bottom panels  show the best simultaneous fit to
the Nobeyama moment $\mu_1$ (beamwidth $5\farcs0$), and the ALMA moments
$\mu_1^\mathrm{blue}$ and $\mu_1^\mathrm{red}$ (beamwidth $0\farcs31$). 
The best fit is obtained for 
an infinite infall radius,  
a systemic velocity $V_\mathrm{sys}=8.29$ \kms, and
a central mass of 0.09 \Msol.
The black dotted line in to bottom panel is not a fit, but the value
of $\mu_1$ predicted by the best-fit model for the ALMA first-order moment.
}
\end{figure}

Regarding the Nobeyama data, the beam obtained for the combined data from both
instruments was $\sim5\farcs0$, and the spectral resolution was 0.108 \kms. The
channel maps of the H$^{13}$CO$^+$ emission were retrieved from Fig.\ 5 of
\citet{Kur13}, and the data cube obtained was resampled with a cell size of
$1''$. The zeroth and first-order moments obtained are shown in Fig. 
\ref{b335_nobeyama_mom}. 
The value of the first-order moment as a function of the angular
distance was obtained by averaging the first-order moment in concentric rings of width
$2''$ centered on the position of the continuum compact source,
\RA{19}{37}{00}{89}, \DEC{+07}{34}{09}{6} \citep{Yen15}, 
the $(0,0)$ position in Figs.\ \ref{b335_nobeyama_mom} and \ref{fig_b335-alma}.
The values obtained are shown in Fig.\ \ref{b335_model} (top panel). The error
bars are the rms dispersion of velocities inside each ring, added quadratically
to the uncertainty due to the finite spectral resolution, the same for all
rings.

The ALMA data had an angular resolution of $\sim0\farcs31$, and the spectral
resolution was 0.166 \kms. 
The first-order moment for all the line, $\mu_1$, is shown in the middle panel
of Fig.\ \ref{fig_b335-alma}. As can be seen, the central blue spot is not
centered on the position of the continuum compact source (plus sign), but
is offset and extends to the southeast. 
The angular resolution of the data corresponds to a linear resolution of 33 au
at a distance of 105 pc. At this small scale, the kinematics can be dominated by
the rotation of the protostellar disk, and infall can be no longer spherically
symmetric, as assumed by the hallmark model.  
In order to check the validity of the model, we also computed the first order
moment separately for the redshifted channels, with velocity higher than the
systemic velocity,  $\mu_1^\mathrm{red}$, 
and for the blueshifted channels (with velocity lower than
$V_\mathrm{sys}$), $\mu_1^\mathrm{blue}$. 
The hallmark model predicts that $\mu_1^\mathrm{red}$ shows a central red spot
and $\mu_1^\mathrm{blue}$ a central blue spot, higher in absolute value than the
first-order moment of the full line. As can be seen in the top and bottom panels of Fig.\
\ref{fig_b335-alma}, this is what is observed for $\mu_1^\mathrm{red}$ and
$\mu_1^\mathrm{blue}$, with quite well centered red and blue spots.
The values of $\mu_1$, $\mu_1^\mathrm{red}$, and $\mu_1^\mathrm{blue}$ as a
function of the angular distance were calculated for concentric rings of
$0\farcs1$ width, centered on the position of the continuum compact source. 
The error bars were calculated in the same way as for the Nobeyama data.
The moments as a function of the angular distance $\theta$, are shown in Fig.\
\ref{b335_model} (bottom panel). 
As can be seen, the three moments follow the same kind of dependence on the
projected distance, and we will see that they can be fitted by power laws of
index $-1/2$, convolved with the beam, as predicted by the hallmark model.  
Thus the gas kinematics for the range of projected distances sampled by ALMA,
from $0\farcs1$ to $0\farcs8$ (10 to 84 au), appears to be dominated by infall. 

The hallmark model was calculated and fitted simultaneously to the Nobeyama
first-order moment $\mu_1$, and to the ALMA  $\mu_1^\mathrm{blue}$  and
$\mu_1^\mathrm{red}$ (but not to $\mu_1$). 
Since the Nobeyama data are very sensitive to the value
adopted for the systemic velocity, we fitted both the value of the systemic
velocity and the central mass.
We tested the infall radius reported by \citet{Kur13}, 2900 au (corrected from
their assumed distance of 150 pc), but a better fit was obtained with a
larger infall radius. 
The best fit was obtained for an infinite infall radius, for a systemic velocity
$V_\mathrm{sys}=8.29$ \kms, and central mass $M_\ast=0.09$ \Msol\ (dashed lines
in the top and bottom panels of Fig.\ \ref{b335_model}). The goodness
of the fit is given by a $\chi^2$ statistic for $\nu=27$ degrees of freedom,
$\chi^2=29.9$, corresponding to a reduced $\chi_r=(\chi^2/\nu)^{1/2}= 1.05$, an
indication that the model fits well the data within the uncertainties.
As an additional check, we computed the predicted value of the first-order
moment for the ALMA data of the full line, $\mu_1$, (dotted line in the bottom
panel of Fig.\ \ref{b335_model}), which is a linear combination of
$\mu_1^\mathrm{blue}$ and  $\mu_1^\mathrm{red}$, i.e.\
$\mu_1= (\mu_0^\mathrm{blue}/\mu_0)\,\mu_1^\mathrm{blue}+
        (\mu_0^\mathrm{red} /\mu_0)\,\mu_1^\mathrm{red}$. 
As can be seen in the figure,
the values predicted for $\mu_1$ match well those observed.

We can conclude that the kinematics of the gas in B335, for the linear 
scales sampled by ALMA and Nobeyama,
from $\sim10$ au to $\sim2500$ au, i.e.\ more than two orders of magnitude, 
can be explained by a simple model of infall onto a central protostar of
$\sim0.1$ \Msol.
This can be considered as an outstanding result of the central-blue-spot
infall hallmark model.

\subsection{LDN 1287}

\object{LDN 1287} (hereafter L1287) is a molecular cloud located at a distance of 929 pc, associated with an
energetic bipolar CO outflow \citep{Yan91}. The source was mapped with a single
dish in \nh\ \citep{Est93, Sep11}. A cluster of mm sources has been
detected at the center of L1287 \citep{Jua18}, one of the mm sources being
associated with VLA 3 \citep{Ang94}, a jet-like cm-continuum source that appears
to be driving the outflow. 

\begin{figure}[htb]
\centering
\resizebox{\hsize}{!}{\includegraphics{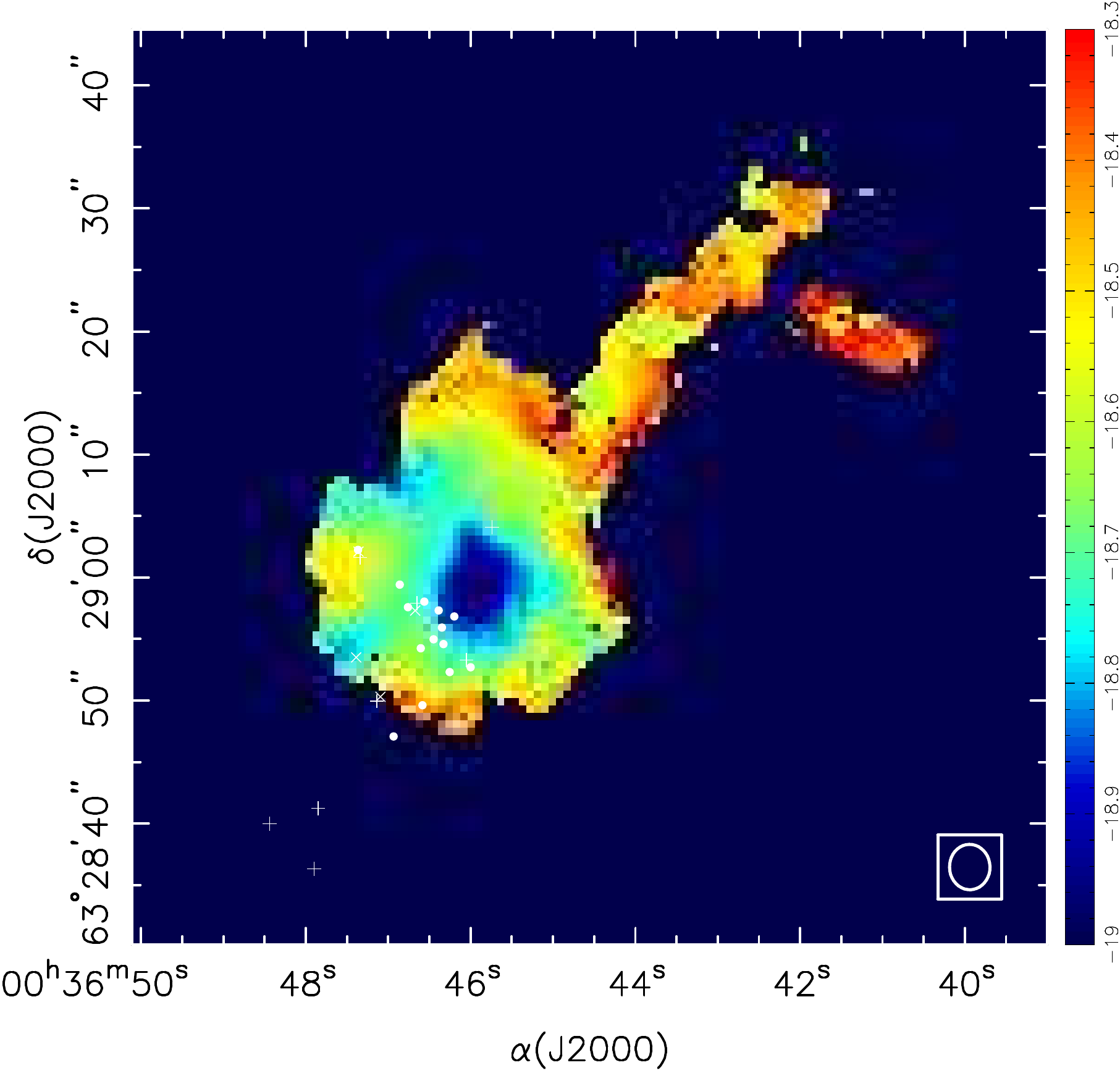}}
\caption{\label{guitar_vlsr}
\nh\ $(1,1)$ line central velocity of the Guitar Core in L1287. The central blue
spot is blueshifted $\sim0.5$ \kms\ with respect to the ambient gas.
The embedded sources in the center are indicated by 
dots \citep[mm, ][]{Jua18}, and 
crosses \citep[cm, ][]{Ang94}.
The synthesized beam is shown in the lower-right corner of the map.
}
\end{figure}

\begin{figure}[htb]
\centering
\resizebox{\hsize}{!}{\includegraphics{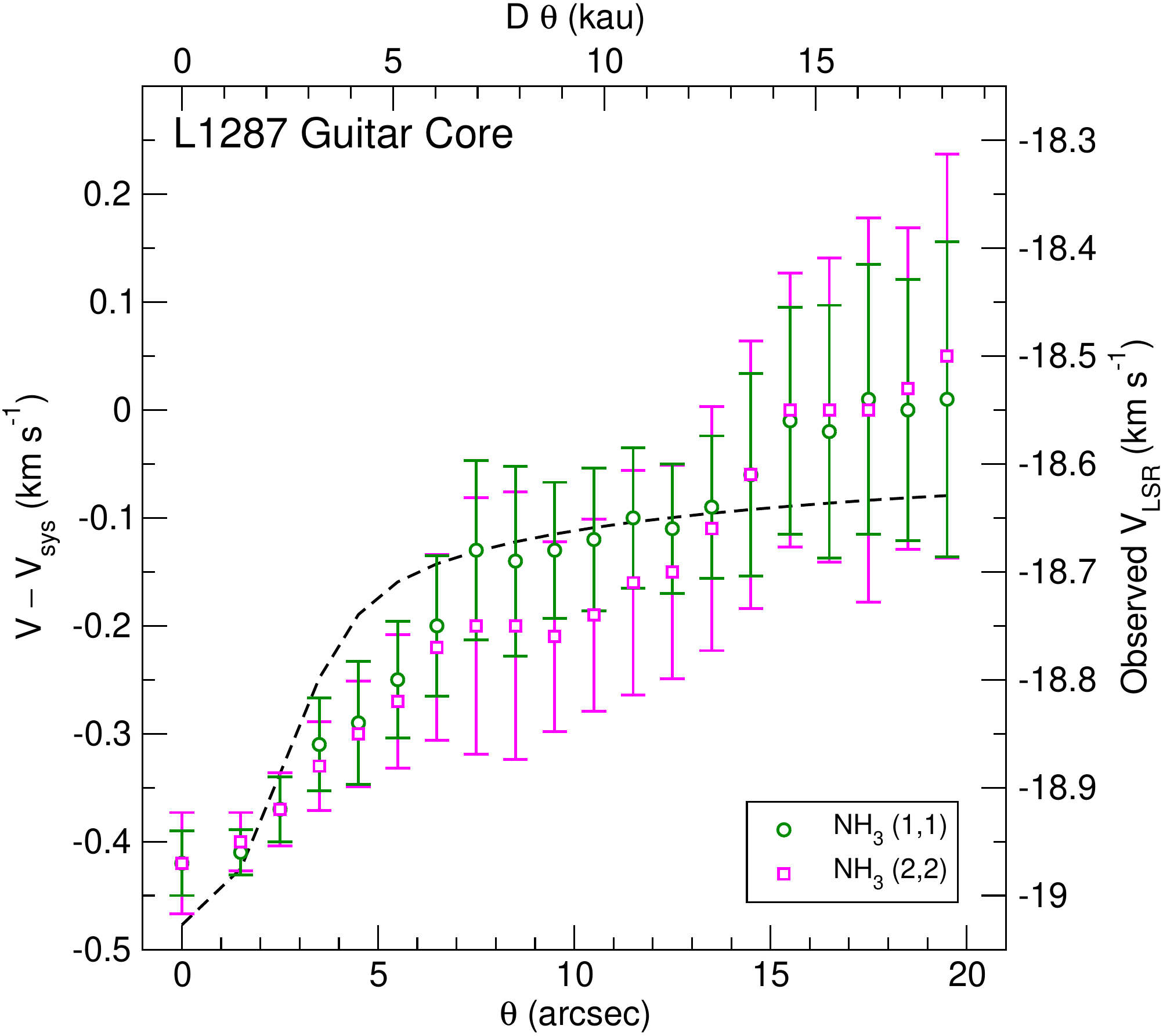}}
\caption{
Same as Fig.\ \ref{g31_mom1} for the Guitar Core in L1287, for the
\nh\ $(1,1)$ (blue circles) and $(2,2)$ lines (red circles). 
The rings used were $1''$ wide, and  
a systemic velocity $V_\mathrm{sys}= -18.55$ km~s$^{-1}$ was adopted.
The best fit was obtained for an infinite
infall radius, and a central mass  $M_\ast=4.4$~\Msol\ (black dashed line). 
\label{guitar_rings}}
\end{figure}

Here we are analyzing VLA observations of the \nh\ $(1,1)$ and $(2,2)$
transitions, which show a complicated structure with a complex kinematics
\citep{Sep18}. The \nh\ lines were analyzed by means of the Hyperfine Structure
Tool (HfS) \citep{Est17}. After careful inspection of the spectra, three
different velocity components were identified, with non-overlapping velocity
ranges (Guitar Core, Blue Filament, and Red Filament). The Guitar Core does not
show any sign of interaction with the embedded young stellar objects (no
increase in linewidth, nor in rotational temperature at the projected position
of the embedded sources). Our results suggest that the Guitar core is a very
young protostellar core. Given the poor velocity resolution of the observations,
the only way to separate the emission of the Guitar Core from that of the
filaments was to fit Gaussian components to the observed spectra. Thus, the
asymmetry of the line was inferred from the shift of the central velocity of the
line fitted. Nevertheless, a compact spot of $\sim10''$ in diameter of
blueshifted velocities appears at the center of the Guitar Core (see Fig.\
\ref{guitar_vlsr}).

The central velocity of the Guitar Core, obtained from the HfS fits, was
averaged in concentric rings $1''$ wide, centered on the emission peak, up to a radius of
$20''$. The velocity profile obtained is shown if Fig.\ \ref{guitar_rings}. The
error bars are the rms dispersion of the velocities averaged in each ring, added
quadratically to the error of the average value of central velocity obtained
from HfS. The average velocity at large distances from the peak (the systemic
velocity of the Guitar Core),
$-18.54$ \kms, 
has been subtracted from the values of the central velocity. The
best fit to the $(1,1)$ and $(2,2)$ data, for a beamwidth of $3\farcs48$, was
obtained for a central mass of 4.8~\Msol\ and a infinite infall radius (dashed
line in Fig.\ \ref{guitar_rings}). The goodness of the fit is indicated by the
value of the $\chi^2$ statistic for $\nu=39$ degrees of freedom (the total
number of rings used in the fit, minus 1), $\chi^2= 42.7$, which gives a reduced
$\chi_r=(\chi^2/\nu)^{1/2}= 1.05$.

\section{Conclusions}

The central-blue-spot infall hallmark \citep{May14} was studied quantitatively,
taking as a basis the work of \citet{Ang87, Ang91}. The assumptions were that 
the line emission was optically thick, 
the gravitational infall motions dominated the kinematics over turbulent and
thermal motions, 
the infall velocity and temperature were power-laws of radius and increase
inwards, with power-law indices of $-1/2$,
and that the Sobolev approximation was valid.
With these assumptions an analytical expression for the first-order moment as a
function of the projected distance was derived, for the cases of infinite
and finite infall radius. 
The effect of a finite angular resolution was also studied, but the convolution
with the beam has to be calculated numerically.
These results were applied to existing data of several star-forming regions
(G31, B335, and L1287),
obtaining good fits to the first-order moment maps, and deriving
values of the central masses onto which the infall is taking place.
The values obtained for the central masses are
70--120 \Msol{} for G31, 
0.1 \Msol{} for B335, and 
4.8 \Msol{} for the Guitar Core of L1287. 

In conclusion, the central-blue-spot infall hallmark appears to be a robust and
reliable indicator of infall.

\begin{acknowledgements} 

This work has been partially supported by the Spanish MINECO grants
AYA2014-57369-C3 and AYA2017-84390-C2 (cofunded with FEDER funds), by
MDM-2014-0369 of ICCUB (Unidad de Excelencia `Mar\'{\i}a de Maeztu'),
and through the ``Center of Excellence Severo Ochoa'' award for the
Instituto de Astrof\'{\i}sica de Andaluc\'{\i}a (SEV-2017-0709).

\end{acknowledgements}

{}

\begin{appendix}

\section{Calculation of the moments for an infinite infall radius and 
arbitrary power-law indices}
\label{ap_infinite}

\subsection{Intensity profile (infinite infall radius)}

Let us assume that the infall velocity and temperature in an infalling molecular
gas core are given by power laws with arbitrary power-law indices, $-\alpha$ and
$-\beta$,
\begin{eqnarray} 
V/V_0 &=& (R/R_0)^{-\alpha}, \nonumber\\ 
T/T_0 &=& (R/R_0)^{-\beta}.
\end{eqnarray}
The development made in \S \ref{Iprofile} for $\alpha=\beta=1/2$  can be
generalized for any positive values of the power-law indices, $\alpha,\beta>0$. 
The projected distance and temperature (Eq.\ \ref{etp}) are now
\begin{eqnarray}\label{eqtp}
p &=& \left(|z/v_z|^{2/(\alpha+1)} - z^2\right)^{1/2}, \nonumber\\
t &=& |v_z/z|^{\beta/(\alpha+1)}.
\end{eqnarray}
The expressions for $z^\ast$, $z_m$, $p_m$ 
(Eq.\ \ref{eq4}) become
\begin{eqnarray}
z^\ast &=& |v_z|^{-1/\alpha},  \nonumber\\ 
z_m &=& (\alpha+1)^{-(\alpha+1)/(2\alpha)} v_z^{-1/\alpha}, \\ 
p_m &=& \alpha^{1/2} z_m = \alpha^{1/2}(\alpha+1)^{-(\alpha+1)/(2\alpha)} v_z^{-1/\alpha}, \nonumber
\end{eqnarray}
and $t(p_m)$ (Eq.\ \ref{eq5}) is now
\begin{equation}
t(p_m)= (\alpha+1)^{\beta/(2\alpha)} v_z^{\beta/\alpha}.
\end{equation}

\subsection{Line profile (infinite infall radius)}

The equations derived in \S \ref{Lprofile} can be generalized as follows.
The temperature and LOS velocity (Eq.\ \ref{eq7}) become
\begin{eqnarray}\label{eqvzt}
v_z &=& \frac{-z}{(p^2+z^2)^{(\alpha+1)/2}}, \nonumber\\ 
t   &=& \frac{1}{(p^2+z^2)^{\beta/2}}.
\end{eqnarray}
The velocity $v_m$ and temperatures $t_1$ and $t_2$ (Eqs.\
\ref{eq8}, \ref{eq9}, \ref{eq11}) are now given by
\begin{eqnarray}\label{eqvm}\label{eqt1}\label{eqt2}
v_m &=& \alpha^{\alpha/2}(\alpha+1)^{-(\alpha+1)/2}\,p^{-\alpha},\nonumber\\
t_1 &=& [\alpha/(\alpha+1)]^{\beta/2}\,p^{-\beta}, \\
t_2 &=& p^{-\beta}. \nonumber
\end{eqnarray}

\subsection{Moments calculation (infinite infall radius)}

From Eq.\ \ref{eqvzt} we can obtain $v_z$ as an explicit function of $t$,
\begin{equation} 
|v_z|= t^{\alpha/\beta}{(1-p^2t^{2/\beta})^{1/2}},
\end{equation}
where the blue-wing profile ($v_z<0$) is obtained for  $t_1<t<t_2$, 
and the red-wing profile ($v_z>0$) for $0<t<t_1$ (see Eq.\ \ref{eqt1}).

In order to calculate the first-order normalized moment $\mu_1(p)$ 
as a function of the projected distance,
\begin{equation}
\mu_1(p)= \frac{\mu'_1}{\mu_0},
\end{equation}
we need to evaluate the integrals 
\begin{eqnarray} 
\mu_0  &=& \int_\mathrm{line}\! t\,dv_z= v_z t - \int\! v_z\,dt, \nonumber\\ 
\mu'_1 &=& \int_\mathrm{line}\! v_zt\,dv_z= 
\frac{1}{2}\,v_z^2 t -\frac{1}{2}\int\! v_z^2\,dt.
\end{eqnarray}

\subsection{Zeroth-order moment (infinite infall radius)}

For the blue wing, the limits of integration are from $(t=t_1, v_z=-v_m)$ to 
$(t=t_2, v_z=0)$. The resulting integral for
$\mu^\mathrm{blue}_0$ is 
\begin{equation}
\mu^\mathrm{blue}_0= 
v_m t_1+ \int_{t_1}^{t_2}\! |v_z|\,dt= 
v_m t_1+ \int_{t_1}^{t_2}\! t^{\alpha/\beta}(1-p^2t^{-2/\beta})^{1/2}\,dt,
\end{equation}
where  $v_m$,  $t_1$, and  $t_2$  have already been defined in Eq.\ \ref{eqt1}. 
The first term is
\begin{equation}
v_m t_1 = \alpha^{(\alpha+\beta)/2} (\alpha+1)^{-(\alpha+\beta+1)/2} p^{-(\alpha+\beta)}.
\end{equation}
The integral of the second term has the same dependence on $p$. This can be seen
with the change of variables $x=pt^{1/\beta}$, resulting in
\begin{equation}
\int_{t_1}^{t_2}\! |v_z|\,dt=
p^{-(\alpha+\beta)}\int_{\![\alpha/(\alpha+1)]^{1/2}}^1 \beta x^{\alpha+\beta-1}\sqrt{1-x^2}\,dx.
\end{equation}
It is useful to define the function $G_0$, 
\begin{equation}\label{eqfprime}
G_0(y) = \int_y^1\! \beta x^{\alpha+\beta-1}\sqrt{1-x^2}\,dx,
\end{equation}
which, in general, is not analytical, but is convergent for $\alpha+\beta>0$, and
the function to integrate is continuum for $x>0$. With this
notation, the zeroth-order moment can be expressed as
\begin{equation}
\mu^\mathrm{blue}_0= B_0\,p^{-(\alpha+\beta)},
\end{equation}
with 
\begin{equation}
B_0= \alpha^{(\alpha+\beta)/2}(\alpha+1)^{-(\alpha+\beta+1)/2}
+G_0\left([\alpha/(\alpha+1)]^{1/2}\right).
\end{equation}

Similarly, for the red wing the limits of integration are from $(t=0, v_z=0)$
to  $(t=t_1, v_z=v_m)$. The resulting integral
for $\mu^\mathrm{red}_0$ is
\begin{equation}
\mu^\mathrm{red}_0= v_m t_1- \int_{0}^{t_1} |v_z|\,dt=
v_m t_1+ \int_{0}^{t_1}\! t^{\alpha/\beta}(1-p^2t^{-2/\beta})^{1/2}\,dt.
\end{equation}
Using the same notation used for the blue wing, we can obtain
\begin{equation}
\mu^\mathrm{red}_0= R_0\,p^{-(\alpha+\beta)},
\end{equation}
with 
\begin{equation}
R_0= B_0-G_0(0),
\end{equation}
where
\begin{equation}
G_0(0) = \int_0^1\! \beta x^{\alpha+\beta-1}\sqrt{1-x^2}\,dx,
\end{equation}
is analytical for $\alpha+\beta>0$, and its value is
\begin{equation}
G_0(0)= \frac{\beta \sqrt{\pi}}{4} 
\frac{\Gamma\left[(\alpha+\beta)/2\right]}{\Gamma\left[(\alpha+\beta+3)/2\right]},
\end{equation}
where $\Gamma$ is the Gamma function.

Thus, the expression for the total zeroth-order moment is
\begin{equation}
\mu_0 =(B_0+R_0)\,p^{-(\alpha+\beta)} = H_0\,p^{-(\alpha+\beta)},
\end{equation}
with
\begin{eqnarray}
H_0 &=& 2B_0-G_0(0) =  2\alpha^{(\alpha+\beta)/2}(\alpha+1)^{-(\alpha+\beta+1)/2} -\nonumber\\
&&-{}\frac{\beta \sqrt{\pi}}{4} 
\frac{\Gamma\left[(\alpha+\beta)/2\right]}{\Gamma\left[(\alpha+\beta+3)/2\right]}
+2G_0\left([\alpha/(\alpha+1)]^{1/2}\right).
\end{eqnarray}

\subsection{First-order moment (infinite infall radius)}

For the blue wing, using the same limits of integration as for the
zeroth-order moment, we have (see Fig.\ \ref{flineprofile})
\begin{equation}
{\mu'_1}^\mathrm{blue}= 
-\frac{1}{2}v_m^2 t_1-\frac{1}{2}\int_{t_1}^{t_2}\! v_z^2\,dt.
\end{equation}
The first term is
\begin{equation}
\frac{1}{2}v_m^2= 
\frac{1}{2} \alpha^{\alpha+\beta/2} (\alpha+1)^{-(\alpha+\beta/2+1)} p^{-(2\alpha+\beta)},
\end{equation}
and the integral can be evaluated using the same change of variables as for the
zeroth-order moment, obtaining
\begin{equation}
\frac{1}{2}\int_0^{t_2}\! v_z^2\,dt=
G_1\left([\alpha/(\alpha+1)]^{1/2}\right) p^{-(2\alpha+\beta)},
\end{equation}
where we defined the function $G_1$ as
\begin{equation}\label{eqgprime}
G_1(y) = \int_y^1\! \frac{\beta}{2} x^{2\alpha+\beta-1}(1-x^2)\,dx,
\end{equation}
which is analytical, and convergent for $2\alpha+\beta>0$, 
\begin{equation}
G_1(y) = \frac{\beta}{(2\alpha+\beta)(2\alpha+\beta+2)} - 
\frac{\beta y^{2\alpha+\beta}}{2}\left[\frac{1}{2\alpha+\beta}-\frac{y^2}{2\alpha+\beta+2}\right].
\end{equation}
Thus, the unnormalized first-order moment of the blue wing can be
expressed as
\begin{equation}
{\mu'_1}^\mathrm{blue}= B_1 p^{-(2\alpha+\beta)},
\end{equation}
with
\begin{equation}
B_1= -\frac{1}{2} \alpha^{\alpha+\beta/2} (\alpha+1)^{-(\alpha+\beta/2+1)}
-G_1\left([\alpha/(\alpha+1)]^{1/2}\right).
\end{equation}
Similarly, for the red wing we have
\begin{equation}
{\mu'_1}^\mathrm{red}= 
\frac{1}{2}v_m^2 t_1-\frac{1}{2}\int_{0}^{t_1}\! v_z^2\,dt,
\end{equation}
that can be expressed as
\begin{equation}
{\mu'_1}^\mathrm{red}= R_1 p^{-(2\alpha+\beta)},
\end{equation}
with
\begin{equation}
R_1= 
\frac{1}{2} \alpha^{\alpha+\beta/2} (\alpha+1)^{-(\alpha+\beta/2+1)}
+G_1\left([\alpha/(\alpha+1)]^{1/2}\right)-G_1(0).
\end{equation}
Note that $B_1$ and $R_1$ cancel each other partially, so that
\begin{equation}
B_1+R_1= -G_1(0)= -\frac{\beta}{(2\alpha+\beta)(2\alpha+\beta+2)}.
\end{equation}
Thus, the expression for the total unnormalized first-order moment is
simply
\begin{equation}
\mu'_1= H_1 p^{-(2\alpha+\beta)},
\end{equation}
with
\begin{equation}
H_1 = B_1+R_1= -\frac{\beta}{(2\alpha+\beta)(2\alpha+\beta+2)}.
\end{equation}

The final results, for an infinite infall radius, can be summarized as
follows. 
\begin{eqnarray}\label{eqmu1alpha}
\mu_0  &=&   H_0\,p^{-(\alpha+\beta)}, \nonumber\\
\mu'_1 &=&   H_1\,p^{-(2\alpha+\beta)},          \nonumber\\
\mu_1  &=&   [H_1/H_0]\,p^{-\alpha},    \\
H_0 &=& 2\alpha^{(\alpha+\beta)/2}(\alpha+1)^{-(\alpha+\beta+1)/2}
-\frac{\beta \sqrt{\pi}}{4} 
\frac{\Gamma\left[(\alpha+\beta)/2\right]}{\Gamma\left[(\alpha+\beta+3)/2\right]} + \nonumber\\
&&+2G_0\left([\alpha/(\alpha+1)]^{1/2}\right),\nonumber\\
H_1 &=& -\frac{\beta}{(2\alpha+\beta)(2\alpha+\beta+2)},        \nonumber
\end{eqnarray}
where $\Gamma$ is the Gamma function, and
\begin{equation}
G_0(y) = \int_y^1\! \beta x^{\alpha+\beta-1}\sqrt{1-x^2}\,dx,
\end{equation}
which, in general, has to be evaluated numerically.

\subsection{Particular case (infinite infall radius) for power-law indices 1/2}
\label{ap_calculmu1}

For the particular case of $\alpha=\beta=1/2$, the former expressions are simpler. In
particular, 
the integral $G_0$ is analytical, 
\begin{equation}\label{eqg0}
G_0(y) = \int_y^1\! \frac{1}{2}\sqrt{1-x^2}\,dx =
\frac{\pi}{8}-\frac{y}{4}\sqrt{1-y^2}-\frac{1}{4}\arcsin y,
\end{equation}
with 
\begin{eqnarray}
G_0(0)          &=& \frac{\pi}{8}                   \nonumber\\
G_0\left(1/\sqrt{3}\right) &=& \frac{\pi}{8}-\frac{\sqrt{2}}{12}-
                    \frac{1}{4}\arcsin\frac{1}{\sqrt{3}}, 
\end{eqnarray}
resulting in 
\begin{eqnarray}\label{eqb0}
B_0 &=& \frac{\pi}{8}+\frac{\sqrt{2}}{4}-\frac{1}{4}\arcsin\frac{1}{\sqrt{3}}= 
0.592,\nonumber\\ 
R_0 &=& B_0-\frac{\pi}{8}= 0.200,
\end{eqnarray}
and
\begin{equation}
H_0 = B_0+R_0 =
\frac{\pi}{8}+\frac{\sqrt{2}}{2}-\frac{1}{2}\arcsin\frac{1}{\sqrt{3}}= 0.792.
\end{equation}
For the first-order moment we have
\begin{eqnarray}
G_1(y) &=& \frac{2}{21}-\frac{y^{3/2}}{2}\left(\frac{1}{3}-\frac{y^2}{7}\right), \nonumber\\
G_1\left(1/\sqrt{3}\right) &=& \frac{2-3^{1/4}}{21}, 
\end{eqnarray}
resulting in
\begin{eqnarray}\label{eqb1}
B_1 &=& -\frac{2}{21}-\frac{4}{63}3^{1/4} = -0.179,\nonumber\\ 
R_1 &=& \frac{4}{63}3^{1/4} = 0.084,
\end{eqnarray}
and the value of $H_1$ is
\begin{equation}
H_1 = B_1+R_1= -\frac{2}{21} = -0.095.
\end{equation}
It can be useful to give the normalized first-order moment separately for the
blue and red wings,
\begin{eqnarray}
\mu_1^\mathrm{blue} &=& [B_1/B_0]\,p^{-1/2} = -0.302\,p^{-1/2}, \nonumber\\
\mu_1^\mathrm{red}  &=& [R_1/R_0]\,p^{-1/2} = \phantom{-}0.418\,p^{-1/2},
\end{eqnarray}
which have the same power-law dependence on the projected distance 
as $\mu_1$.

In conclusion, for the total zeroth and first-order moments we have
\begin{eqnarray}
\mu_0  &=&  H_0\,p^{-1}         =  0.792\,p^{-1},   \nonumber\\
\mu'_1 &=&  H_1\,p^{-3/2}       = -0.095\,p^{-3/2}, \\
\mu_1  &=&  [H_1/H_0]\,p^{-1/2} = -0.120\,p^{-1/2}. \nonumber
\end{eqnarray}

\section{Calculation of the moments for a finite infall radius and 
arbitrary power-law indices}
\label{ap_finite}

The critical value of the reduced coordinate $q= p/r_i$, for an arbitrary
value of the power-lax index $\alpha$, is $q=[\alpha/(\alpha+1)]^{1/2}$. 
For values $q<[\alpha/(\alpha+1)]^{1/2}$, only the red-wing emission is affected;  
for $q\ge[\alpha/(\alpha+1)]^{1/2}$ the line becomes symmetric ($\mu_1=0$);
for $q\ge1$ all the wing emission disappears ($\mu_0=\mu_1=0$).

\subsection{Line profile (finite infall radius)}

In order to calculate the moments of the red wing emission we need the values of
$v_a$, $t_a$, and $t_b$ as a function of $p$ and $q$ (see Fig.\
\ref{flineprofile_ri}). For this, we need the
values of $z_a$ and $z_b$, the $z$ coordinate of the two intersections of the
equal-LOS-velocity surface with the line-of-sight.

The distance $z_a$ is obtained readily from $r_i^2=p^2+z_a^2$, which gives
\begin{equation}\label{eza}
z_a= q^{-1} (1-q^2)^{1/2} p,
\end{equation}
and, from Eq.\ \ref{eqt1} we have
\begin{eqnarray}
v_a &=& q^\alpha (1-q^2)^{1/2} p^{-\alpha}, \nonumber\\
t_a &=& q^\beta p^{-\beta}.
\end{eqnarray}

The distance $z_b$ requires more work. The equation of the equal-LOS-velocity
surface of velocity $v_a$ (Eq.\ \ref{eqt1}) can be written as
\begin{equation}
z= v_a (p^2+z^2)^{(\alpha+1)/2}.
\end{equation}
For a given $p$ and $v_a$, $z_b$ is a root of this equation that satisfies
$0<z_b<z_a$. It is useful to use the variable $x=1+(z/p)^2$, 
so that the equation to solve depends only on $q$ and $\alpha$, and becomes
\begin{equation}
x= 1+q^{2\alpha}(1-q^2) x^{\alpha+1}.
\end{equation}
For rational values of $\alpha$ this equation is a polynomial in $x$.  However,
in general, the root has to be found numerically. An iterative algorithm that
gives the correct root for $q<[\alpha/(\alpha+1)]^{1/2}$ is the following
\begin{equation}\label{eqiterative}
\left\{\begin{array}{l}
x_0     = 1, \\
x_{n+1} = 1+q^{2\alpha} (1-q^2) x_n^{\alpha+1}, \quad n=0, 1, 2, \ldots
\end{array}\right.
\end{equation}
It is useful to introduce the parameter $q'$, which plays a role similar to that
of $q$,
\begin{equation}\label{eqqprime}
q'= \frac{p}{(p^2+z_b^2)^{1/2}}= x^{-1/2},
\end{equation}
so that the temperature $t_b$, corresponding to $z_b$, can be expressed as
\begin{equation}
t_b= {q'}^{\beta} p^{-\beta}.
\end{equation}
It can be easily shown that for $q<[\alpha/(\alpha+1)]^{1/2}$, the parameter $q'$ is
always bounded by $q<q'<1$, and that for $q=[\alpha/(\alpha+1)]^{1/2}$, $q'=q$
(see Fig.\ \ref{fqprime}). 

\subsection{Zeroth-order moment (finite infall radius)}

The zeroth-order moment of the blue wing, $\mu_0^\mathrm{blue}$, does not depend on $q$,
and we have already seen that it can be expressed as
\begin{equation}
\mu_0^\mathrm{blue} = B_0\, p^{-(\alpha+\beta)}.
\end{equation}
The zeroth-order moment of the red wing, $\mu_0^\mathrm{red}$, can be
calculated as
\begin{equation}
\mu_0^\mathrm{red} = 
\mu_0^\mathrm{blue} + v_a(t_b-t_a) -\int_{t_a}^{t_b}\! v_z\,dt.
\end{equation}
Using the expressions obtained for $v_a$, $t_a$, and $t_b$, $\mu_0^\mathrm{red}$
can be expressed as
\begin{equation}
\mu_0^\mathrm{red} = R_0(q)\, p^{-(\alpha+\beta)},
\end{equation}
where $R_0(q)$ is
\begin{equation}
R_0(q) = B_0 +q^\alpha (1-q^2)^{1/2} ({q'}^\beta-q^\beta)-G_0(q)+G_0(q'),
\end{equation}
where $q'$ is given by Eq.\ \ref{eqqprime}, and $G_0$ by Eq.\ \ref{eqfprime}.

Finally, the total zeroth-order moment can be written as
\begin{equation}
\mu_0 = [B_0+R_0(q)]\,p^{-(\alpha+\beta)}= H_0(q)\,p^{-(\alpha+\beta)} .
\end{equation}
with
\begin{equation}\label{eqh0}
H_0(q) = 2B_0 + q^\alpha (1-q^2)^{1/2} ({q'}^\beta-q^\beta)-G_0(q)+G_0(q').
\end{equation}

\subsection{First-order moment (finite infall radius)}

The first-order moment for the blue wing does not depend on $q$, and we have already seen
that
\begin{equation}
{\mu'_1}^\mathrm{blue} = B_1 p^{-(2\alpha+\beta)}.
\end{equation}
The first-order moment of the red wing is the opposite of ${\mu'_1}^\mathrm{blue}$, except
for a deficit of red-wing emission (see Fig.\ \ref{flineprofile_ri}), so that
\begin{equation}
{\mu'_1}^\mathrm{red} = {-\mu'_1}^\mathrm{blue}
+\frac{1}{2}v_a^2(t_b-t_a)-\frac{1}{2}\int_{t_a}^{t_b}\! v_z^2\,dt.
\end{equation}
Taking into account the expressions of $v_a$, $t_a$, and $t_b$, the
first-order moment of the red wing can be written as
\begin{equation}
{\mu'_1}^\mathrm{red} = R_1(q)\,p^{-(2\alpha+\beta)},
\end{equation}
where $R_1(q)$ is
\begin{equation}\label{eqh1}
R_1(q)= -B_1
+\frac{q^{2\alpha}}{2}(1-q^2)({q'}^\beta-q^\beta)-G_1(q)+G_1(q'),
\end{equation}
where $q'$ is given by Eq.\ \ref{eqqprime}, and $G_1$ by Eq.\ \ref{eqgprime}.
Thus, the total unnormalized first-order moment is
\begin{equation}
\mu'_1 = [B_1+R_1(q)]\,p^{-(2\alpha+\beta)} = H_1(q)\,p^{-(2\alpha+\beta)}.
\end{equation}
with $H_1$ given by
\begin{equation}
H_1(q)=  \frac{q^{2\alpha}}{2}(1-q^2)({q'}^\beta-q^\beta)-G_1(q)+G_1(q').
\end{equation}

The normalized first order moments for the blue and red wings separately are
given by
\begin{eqnarray}
\mu_1^\mathrm{blue} &=& [B_1/B_0]\,p^{-\alpha}, \nonumber\\
\mu_1^\mathrm{red}  &=& [R_1(q)/R_0(q)]\,p^{-\alpha},
\end{eqnarray}
where $B_0$, $B_1$, $R_0(q)$, and $R_1(q)$ have already been given.
Finally, the expression obtained for the normalized first-order moment of the
whole line, 
$\mu_1=\mu'_1/\mu_0$, is
\begin{equation}
\mu_1 = \frac{H_1(q)}{H_0(q)}\,p^{-\alpha}.
\end{equation}

\subsection{Final results (finite infall radius)}

Assuming infall velocity and temperature that are power laws of the radius,
\begin{eqnarray}
v &=& (R/R_0)^{-\alpha}, \nonumber\\
t &=& (R/R_0)^{-\beta},
\end{eqnarray}
we have, for a finite infall radius $r_i$, with $q=p/r_i$,
\begin{eqnarray}\label{eqmu1_ri}
&&\hspace{-0.8cm}\left.\begin{array}{l}
\mu_0(p)  = H_0(q)\,p^{-(\alpha+\beta)} \\
\mu'_1(p) = H_1(q)\,p^{-(2\alpha+\beta)} \\
\mu_1(p)  = [H_1(q)/H_0(q)]\,p^{-\alpha}
\end{array}\right\}  \quad (0\leq q<[\alpha/(\alpha+1)]^{1/2}), \nonumber\\
&&\hspace{-0.8cm}\left.\begin{array}{l} 
\mu_0(p) = 2B_0\,p^{-(\alpha+\beta)} \\
\mu'_1(p) = \mu_1(p)=0
\end{array}\right\}  \quad ([\alpha/(\alpha+1)]^{1/2}\leq q<1), \\
&&\hspace{-0.8cm}\begin{array}{l}
\mu_0(p) = \mu'_1(p) = \mu_1(p) =0 
\end{array} \quad (q\ge1) \nonumber
\end{eqnarray}
with $H_0(q)$ and $H_1(q)$ given by
\begin{eqnarray}\label{eqh0h1}
H_0(q) &=& 2B_0+q^\alpha(1-q^2)^{1/2}({q'}^{1/2}-q^{1/2})-G_0(q)+G_0(q'),\nonumber \\
H_1(q) &=& \frac{q^{2\alpha}}{2}(1-q^2)({q'}^{1/2}-q^{1/2})-G_1(q)+G_1(q'),
\end{eqnarray}
where $q'$ is an auxiliary parameter 
\begin{equation}
q'= x^{-1/2},\\
\end{equation}
and $x$ is the root, satisfying $0<x<q^{-2}$, of the equation
\begin{equation}
x= 1+q^{2\alpha}(1-q^2) x^{\alpha+1},
\end{equation}
to be solved numerically (see Eq.\ \ref{eqiterative}),
and
\begin{eqnarray}
G_0(y) &=& \int_y^1\! \beta x^{\alpha+\beta-1}\sqrt{1-x^2}\,dx \quad\mbox{(to be done
            numerically),} \nonumber\\
G_1(y) &=& \frac{\beta}{(2\alpha+\beta)(2\alpha+\beta+2)}- \nonumber\\
       &-& \frac{\beta y^{2\alpha+\beta}}{2}
           \left[\frac{1}{2\alpha+\beta}-\frac{y^2}{2\alpha+\beta+2}\right], 
           \\
B_0    &=& \alpha^{(\alpha+\beta)/2} (\alpha+1)^{-(\alpha+\beta+1)/2}
          +G_0\left([\alpha/(\alpha+1)]^{1/2}\right).\nonumber
\end{eqnarray}

\subsection{Particular case (finite infall radius) for power-law indices 1/2}

\begin{figure}[htb]
\centering
\resizebox{\hsize}{!}{\includegraphics{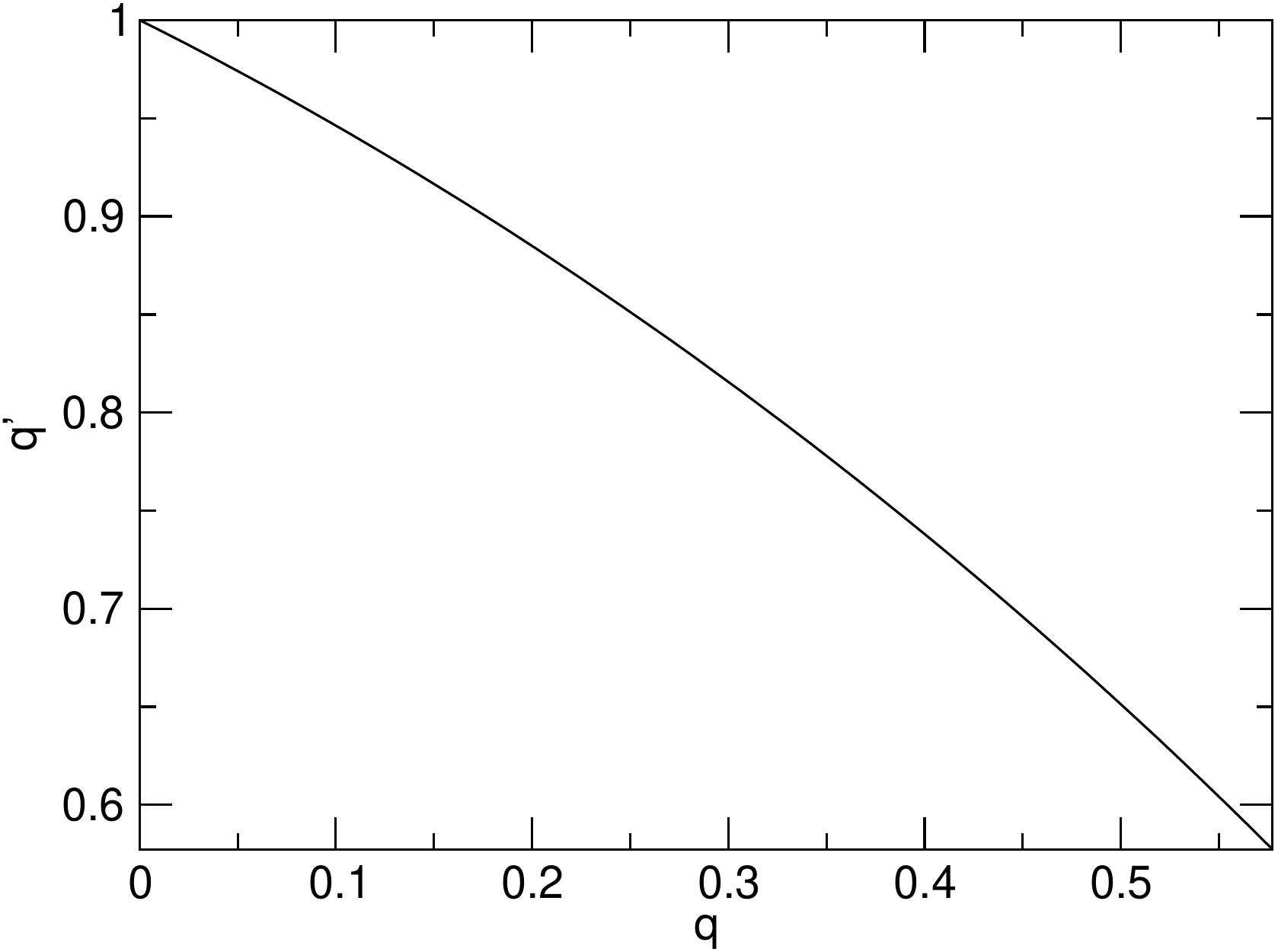}}
\caption{\label{fqprime}
Plot of $q'$ as a function of $q$ (case $\alpha=\beta=1/2$), for $0\le q\le 1/\sqrt{3}$. 
Note that $q\le 1/\sqrt{3}\le q'\le 1$, and that $q'=q$ for $q=1/\sqrt{3}$.
}
\end{figure}

For the particular case of $\alpha=\beta=1/2$, the former expressions are simpler. In
particular, as already seen, the integral $G_0$ is analytical, 
and the equation in $x$ to find $q'=x^{-1/2}$ is a 3rd degree polynomial,
\begin{equation}
P_3(x)= q^2(1-q^2)^2 x^3-(x-1)^2= 0.
\end{equation}
Since we already know a root corresponding to $z_a$, $x= q^{-2}$, the polynomial
is divisible by $(x-q^{-2})$,
\begin{equation}
P_3(x)= q^2(x-q^{-2})[(1-q^2)^2x^2-(2-q^2)x+1]= 0,
\end{equation}
and thus the root $x$ (and $q'$) can be found analytically (see Fig.\ 
\ref{fqprime}),
\begin{equation}\label{eqqp}
q'= \frac{1-q^2}{\left[1-\dfrac{q^2}{2}+\dfrac{q}{2}(4-3q^2)^{1/2}\right]^{1/2}}.
\end{equation}

\begin{figure}[htb]
\centering
\resizebox{\hsize}{!}{\includegraphics{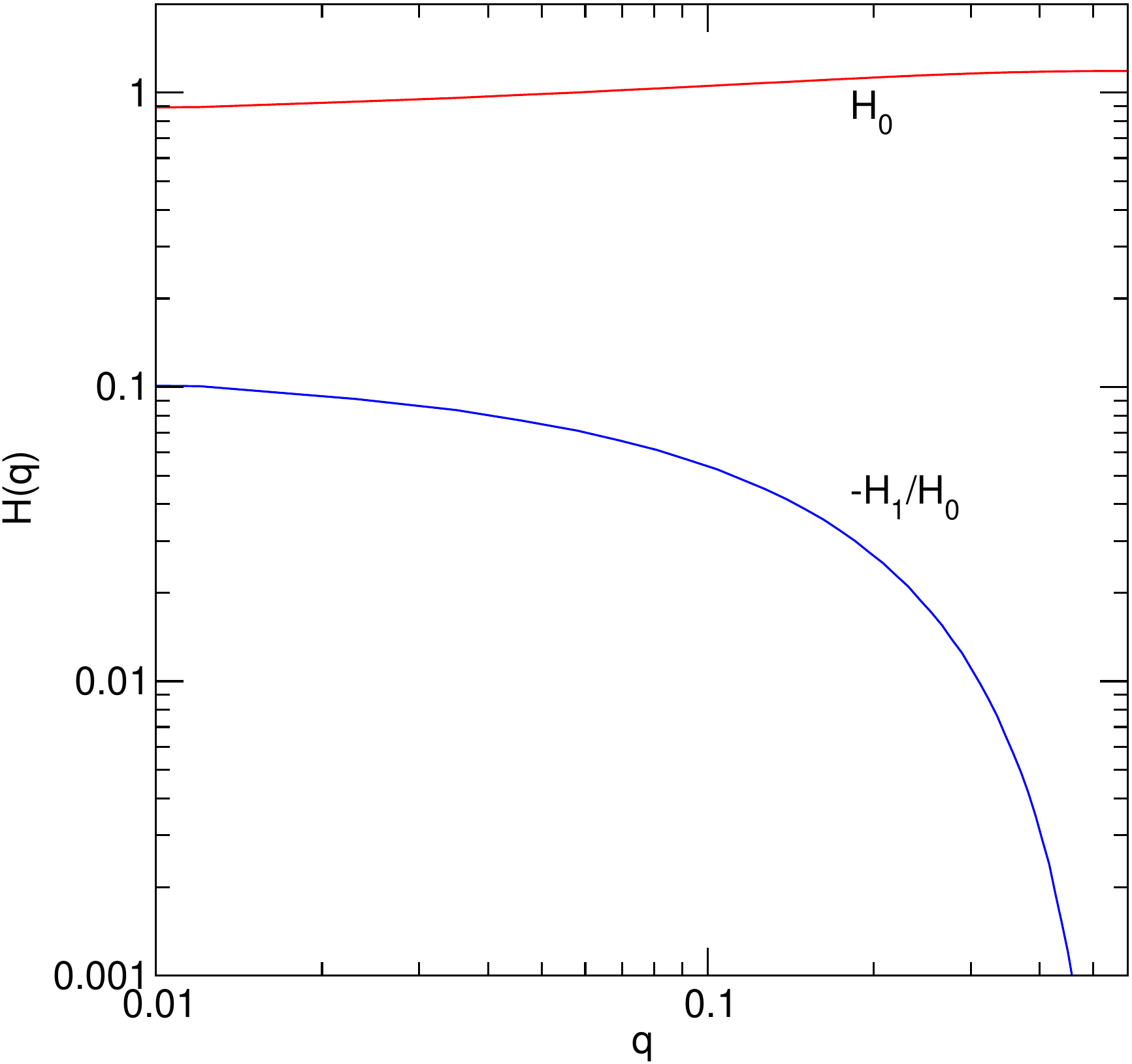}}
\caption{\label{fh01_ri}
Log-log plot of $H_0$ and $-H_1/H_0$ as a function of $q$ (case
$\alpha=\beta=1/2$), for  $0<q<1/\sqrt{3}$.
}
\end{figure}

Thus, the final results for this particular case are
\begin{eqnarray}
&&\hspace{-0.8cm}\left.\begin{array}{l}
\mu_0(p)  = H_0(q)\,p^{-1} \\
\mu'_1(p) = H_1(q)\,p^{-3/2} \\
\mu_1(p)  = [H_1(q)/H_0(q)]\,p^{-1/2}
\end{array}\right\}  \quad (0\leq q<1/\sqrt{3}), \nonumber\\
&&\hspace{-0.8cm}\left.\begin{array}{l} 
\mu_0(p) = 2B_0\,p^{-1} \\
\mu'_1(p) = \mu_1(p)=0
\end{array}\right\}  \quad (1/\sqrt{3}\leq q<1), \\
&&\hspace{-0.8cm}\begin{array}{l}
\mu_0(p) = \mu'_1(p) = \mu_1(p) =0 
\end{array} \quad (q\ge1) \nonumber
\end{eqnarray}
where 
$B_0$ is given by Eq.\ \ref{eqb0},
$H_0$ and $H_1$ (see Fig. \ref{fh01_ri}) are given by 
\begin{eqnarray}
H_0(q) &=& 2B_0+[q(1-q^2)]^{1/2}({q'}^{1/2}-q^{1/2})-G_0(q)+G_0(q'),\nonumber \\
H_1(q) &=& \frac{q}{2}(1-q^2)({q'}^{1/2}-q^{1/2})-G_1(q)+G_1(q'),
\end{eqnarray}
$q'$ is given by Eq. \ref{eqqp}, 
$G_0$ is given by Eq.\ \ref{eqg0}, 
and $G_1$ results in 
\begin{equation}
G_1(y) = \frac{2}{21}-\frac{y^{3/2}}{2}\left(\frac{1}{3}-\frac{y^2}{7}\right).  
\end{equation}

\subsection{Finite angular resolution (finite infall radius) for power-law
indices 1/2}

Although the first-order moment for a finite infall radius $r_i$, 
observed with a finite angular resolution, has to
be calculated numerically, we can study its value at the origin, for $p=0$.
The convolution of
the moments $\mu_0(p;r_i)$ and $\mu'_1(p;r_i)$
with a Gaussian beam of unit area and half-power beamwidth $b$,
\begin{equation}
B(p)= \frac{4\ln2}{\pi b^2}\,e^{-4\ln2\,p^2/b^2},
\end{equation}
for $p=0$ is simply
\begin{eqnarray}
\mu_0(0;b,r_i)  &=& \int_0^\infty H_0(q) p^{-1}   B(p)\, 2\pi p\,dp, \nonumber\\
\mu'_1(0;b,r_i) &=& \int_0^\infty H_1(q) p^{-3/2} B(p)\, 2\pi p\,dp.
\end{eqnarray}
It is useful to use the variable $s\equiv r_i/b$, so that both moments can be
expressed as
\begin{eqnarray}
\mu_0(0;b,r_i)  &=& 8\ln2\, J_0(s)\, b^{-1}, \nonumber\\
\mu'_1(0;b,r_i) &=& 8\ln2\, J_1(s)\, b^{-3/2},
\end{eqnarray}
where
\begin{eqnarray}\label{eqJ}
J_0(s) &=& \int_0^\infty H_0(x/s)\,          e^{-4\ln2\,x^2} dx, \nonumber\\
J_1(s) &=& \int_0^\infty H_1(x/s)\, x^{-1/2} e^{-4\ln2\,x^2} dx.
\end{eqnarray}
The normalized first-order moment at the origin will be
\begin{equation}\label{eqJ1J0}
\mu_1(0;b,r_i)= \frac{J_1(s)}{J_0(s)}\, b^{-1/2}.
\end{equation}
The asymptotic values of $\mu_1(0;b,r_i)$ for $r_i\to\infty$ (equivalent to
$s\gg1$) must coincide  with the expression derived with infinite infall radius
(Eq.\ \ref{eqmu1_0}). Effectively,
\begin{eqnarray}
J_0(s\gg1) &\simeq& H_0(0)\int_0^\infty e^{-4\ln2\,x^2} dx =
H_0\,\frac{\pi^{1/2}}{4(\ln2)^{1/2}}, \\
J_1(s\gg1) &\simeq& H_1(0)\int_0^\infty x^{-1/2} e^{-4\ln2\,x^2} dx =
H_1\,\frac{\Gamma(1/4)}{2^{3/2}(\ln2)^{1/4}},\nonumber
\end{eqnarray}
so that 
\begin{eqnarray}
\mu_1(0;b,r_i\!\gg\!b) 
\!\!\!\!&\simeq&\!\!\!\!
\frac{H_1}{H_0}\,\Gamma(1/4)\,(\ln2)^{1/4}(2/\pi)^{1/2}\, b^{-1/2}\\
\!\!\!\!&=&\!\!\!\! -0.317\, b^{-1/2},\nonumber
\end{eqnarray}
as expected. Let us now evaluate the asymptotic expression for $\mu_1(0;b,r_i)$
for $b\to\infty$ (equivalent to $s\ll1$). Taking into account that 
$H_0(q)=0$ for $q>1$ and 
$H_1(q)=0$ for $q>1/\sqrt{3}$, 
\begin{eqnarray}
J_0(s\ll1) \!\!\!\!&\simeq&\!\!\!\!\!
\int_0^\infty\!\!\!\! H_0(x/s)\, dx =\! 
\int_0^1\!\! H_0(q) s\,dq= C_0\,s, \\
J_1(s\ll1) \!\!\!\!&\simeq&\!\!\!\!\! 
\int_0^\infty\!\!\!\! H_1(x/s)\,x^{-1/2}dx =\!
\int_0^{1/\!\sqrt{3}}\!\!\!\!\!\!\!\!\!\! H_1(q)\, q^{-1/2} s^{1/2} dq= 
C_1 s^{1/2},\nonumber
\end{eqnarray}
where 
$C_0=\int_0^1H_0(q)\,dq$ and 
$C_1=\int_0^{1/\!\sqrt{3}}H_1(q)\,q^{-1/2}dq$ 
are constants to be evaluated numerically, resulting in
$C_1/C_0\simeq-0.060$. Thus,
\begin{eqnarray}
\mu_1(0;b\gg r_i,r_i) &=& \frac{J_1(s\ll1)}{J_0(s\ll1)}\,b^{-1/2}\simeq
\frac{C_1}{C_0}s^{-1/2} b^{-1/2}\\
&\simeq& -0.060\,r_i^{-1/2}.\nonumber
\end{eqnarray}

\section{Finite spectral resolution and first-order moment of a line}
\label{ap_spectralresolution}

Let $I(v)$ be the intensity of a line as a function of radial velocity.
We are interested in calculating the first-order moment of the line profile,
\begin{equation}
\mu_1= \frac{\mu'_1}{\mu_0},
\end{equation}
where
\begin{equation}
\mu_0  = \int_{-\infty}^{+\infty}\!\! I(v)\,dv, \qquad
\mu'_1 = \int_{-\infty}^{+\infty}\!\! v\,I(v)\,dv. 
\end{equation}

We will make use of the relationship between the zeroth and first-order
moments of a function
$f$ and the values of its Fourier transform $\F{f}$ and its derivative $\Fp{f}$ 
at the origin \citep[see][chapter 8]{Bra00}
\begin{equation}
\mu_0= \F{f}(0), \quad
\mu'_1= \frac{\Fp{f}(0)}{-2\pi i}.
\end{equation}

Let us assume that the finite spectral resolution of the spectrometer
can be represented by the convolution of the real line profile with an
instrumental response function $W(v)$,
\begin{equation}
I^\ast(v)= (I \ast W)(v),
\end{equation}
and that the instrumental response is normalized to unit area, and
is symmetric with respect to velocity or, more precisely, with zero first-order
moment,
\begin{equation}
\int_{-\infty}^{+\infty}\!\!    W(v)\,dv=1, \quad
\int_{-\infty}^{+\infty}\!\! v\,W(v)\,dv=0.
\end{equation}
In the Fourier domain we have
\begin{equation}
\F{W}(0)=1, \quad \frac{\Fp{W}(0)}{-2\pi i}=0.
\end{equation}

The zeroth-order moment calculated from the observed profiles will be
\begin{equation}
\mu_0^\ast=  \F{I \ast W}(0) = \F{I}(0)\,\F{W}(0)= \F{I}(0)= \mu_0,
\end{equation}
where we used that the Fourier transform of a convolution is the product of
Fourier transforms \citep{Bra00}. Thus, the zeroth-order moment is
independent of the spectral resolution.

Let us consider now the unnormalized first-order moment,
\begin{equation} 
{\mu'_1}^\ast= 
\int_{-\infty}^{+\infty}\!\! v\,I^\ast(v)\,dv.
\end{equation}
By using the relations with the derivative of the Fourier transform, we obtain
\begin{eqnarray}
{\mu'_1}^\ast &=& 
\frac{\Fp{I \ast W}(0)}{-2\pi i} = 
\frac{(\F{I}\,\F{W})'(0)}{-2\pi i} =
\frac{\Fp{I}(0)}{-2\pi i}\F{W}(0) + \nonumber\\&&
\F{I}(0)\frac{\Fp{W}(0)}{-2\pi i}=
\frac{\Fp{I}(0)}{-2\pi i} = \mu'_1.
\end{eqnarray}
Thus, $\mu'_1$ is independent of the spectral resolution, and so is the
normalized first-order moment, $\mu_1=\mu'_1/\mu_0$.

In conclusion. the spectral resolution does not affect the value of the 
first-order moment of a line, as long as the spectrometer response can
be described as a convolution with an instrumental response symmetric with
respect to velocity.

\section{2-D Convolution of a power-law function with a Gaussian beam}
\label{ap_convolution}

Let $F(p)$ be a 2-D power-law function of the radial distance
\begin{equation}
F(p)= A\,p^{-m},
\end{equation}
that has to be convolved with a
Gaussian beam of unit area and half-power beamwidth $b$,
\begin{equation}\label{ebeam}
B(p)= \frac{4\ln2}{\pi b^2}\,e^{-4\ln2\,p^2/b^2}.
\end{equation}
The 2-D convolution
\begin{equation}
F_b= F \ast B,
\end{equation}
will have a different behavior for $p\gg b$ and $p\ll b$. On the one hand,
for $p\gg b$ the convolution with the Gaussian beam will not modify
noticeably the power-law function and the result will not depend on $b$,
\begin{equation}
F_b(p)\simeq F(p) \quad (p\gg b).
\end{equation}
On the other hand, for $p\ll b$, $F_b$ will not depend much on $p$
and will be approximately constant, 
\begin{equation}
F_b(p)\simeq F_0 \quad (p\ll b),
\end{equation}
with $F_0$ given by
\begin{equation}
F_0= \int_0^\infty \!\! F(p)\,B(p)\,2\pi p\,dp=
2\pi A\frac{4\ln2}{\pi b^2} 
\int_0^\infty \!\! p^{1-m}\,e^{-4\ln2\,p^2/b^2}dp.
\end{equation}
The integral can be evaluated analytically and the result is
\begin{equation}
F_0= (4\ln2)^{m/2}\Gamma(1-m/2)\,A\,b^{-m} .
\end{equation}
For the cases of interest, we have
\begin{equation}\label{eqF0}
F_0=\left\{\begin{array}{ll}
1.5813\,A\,b^{-1/2} & (m=1/2)\\
2.9513\,A\,b^{-1}   & (m=1)\\
7.7901\,A\,b^{-3/2} & (m=3/2)
\end{array}\right..
\end{equation}
The characteristic radius that separates the two regions of $F_b$ can
be estimated as the intersection of the two asymptotic values, i.e.\
$F(p_c)=F_0$, resulting in
\begin{equation}
p_c= \frac{b}{(4\ln2)^{1/2}[\Gamma(1-m/2)]^{1/m}}.
\end{equation}
For the cases of interest, we have
\begin{equation}\label{eqFs}
p_c=\left\{\begin{array}{ll}
0.3999\,b & (m=1/2)\\
0.3388\,b & (m=1)\\
0.2545\,b & (m=3/2)
\end{array}\right..
\end{equation}

\end{appendix}

\listofobjects

\end{document}